\documentclass[twocolumn,usenatbib]{mnras}
\usepackage{graphicx}
\usepackage{lscape}
\usepackage{bookmark}
\usepackage{threeparttablex}
\usepackage{tabularx}
\usepackage[T1]{fontenc}
\usepackage{ae,aecompl}
\usepackage{adjustbox}
\usepackage{subcaption}
\usepackage{savesym}
\usepackage{amsmath}
\savesymbol{iint}
\usepackage{wasysym}
\usepackage{graphicx}	
\usepackage{amssymb}	
\usepackage{etoolbox}
\usepackage{threeparttable}
\usepackage{environ}
\usepackage{hyperref}
\usepackage{rotating}
\usepackage{adjustbox}

\usepackage{soul}


\title[Study of BMP stars using UVIT/AstroSat]{Field blue straggler stars: Discovery of white dwarf companions to blue metal-poor stars using UVIT/AstroSat}
\author[A. Panthi et al.]{
Anju Panthi,$^{1}$\thanks{p20190413@pilani.bits-pilani.ac.in}
Annapurni Subramaniam,$^{2}$
Kaushar Vaidya,$^{1}$ 
Vikrant Jadhav,$^{3}$ 
Sharmila Rani,$^{2}$
\newauthor
Sivarani Thirupathi,$^{2}$
Sindhu Pandey,$^{4}$
\\
$^{1}$ Department of Physics, Birla Institue of Technology and Science, Pilani, Rajasthan-333031, India\\
$^{2}$ Indian Institute of Astrophysics, Sarjapur Road, Koramangala, Bangalore, India\\
$^{3}$ Helmholtz-Institut für Strahlen-und Kernphysik, Universität Bonn, Nussallee 14-16, D-53115 Bonn, Germany\\
$^{4}$ Aryabhatta Research Institute of Observational Sciences, Manora Peak, Nainital, India\\
}

\begin{document}
\label{firstpage}
\pagerange{\pageref{firstpage}--\pageref{lastpage}}
\maketitle

\begin{abstract}
Blue metal-poor (BMP) stars are the main-sequence stars that appear bluer and more luminous than normal turn-off stars of metal-poor globular clusters. They are believed to be either field blue straggler stars (FBSS) formed via post-mass transfer mechanism or accreted from dwarf satellite galaxies of the Milky Way. A significant fraction of BMP stars are discovered to be potential binaries. We observed 27 BMP stars using UVIT/\textit{AstroSat} in two FUV filters, F148W and F169M. We report the discovery of white dwarf (WD) companions of 12 BMP stars for the first time. The WD companions have estimated temperatures T$_{eff}$ $\sim$10500 $-$ 18250 K, and masses 0.17 M$_{\odot}$ $-$ 0.8 M$_{\odot}$. Based on [Fe/H] and space velocity, we group the 12 BMP/FBSS stars as the thick disk (5) and halo (5), whereas two stars appear to be in-between. All the 5 thick disk BMP/FBSS have extremely low-mass (M $<$ 0.2 M$_{\odot}$) WDs as companions, whereas the 5 halo BMP/FBSS have low (0.2 M$_{\odot}$ $<$ M $<$ 0.4 M$_{\odot}$), normal (0.4 M$_{\odot}$ $<$ M $<$ 0.6M$_{\odot}$), and high mass (M $>$ 0.6 M$_{\odot}$) WD companions. Our analysis suggests that at least $\sim$44 $\%$ of BMP stars are FBSS, and these stars hold the key to understand the details of mass transfer, binary properties, and chemical enrichment among the FBSS.
\end{abstract}


\section{Introduction} \label{Introduction}
Blue straggler stars (BSS) are the enigmatic stellar populations that were first discovered by \citet{sandage1953color} in the globular cluster (GC) M3, and are brighter and bluer than the main sequence turn-off on the colour-magnitude diagram (CMD). The fact that BSS are more massive than conventional main-sequence (MS) stars \citep{shara1997first, gilliland1998oscillating, de2005spectroscopic, beccari2006population}, implies that they were formed by a process capable of increasing their initial mass. In this context, mass transfer (MT) by Roche lobe overflow in primordial binary systems \citep{mccrea1964extended, 1976ApJ...209..734Z}, merger or MT of an inner binary in a triple system \citep{perets2009triple}, and direct stellar collisions \citep{hills1976stellar, chatterjee2013stellar} are the most plausible mechanisms for their formation. MT mechanism can further be classified into three categories depending on the location of the primary star while transferring mass to the secondary, Case A: when the primary star is on the MS \citep{webbink1976evolution}, Case B: when the primary star is on the red giant branch (RGB) phase \citep{mccrea1964extended}, and Case C: when the primary star is on the asymptotic giant branch (AGB) phase \citep{chen2008binary}. 

Despite being exotic, BSS have been identified in essentially all stellar environments: GCs \citep{sandage1953color}, open clusters (OCs, \citealt{ahumada1995catalogue}), field population of the Milky Way \citep{preston2000these}, and dwarf galaxies \citep{momany2007blue}. However, different formation channels of BSS, as stated above, operate depending on the environment \citep{fusi1992blue}. In the case of GCs, collisions and MT processes are expected to act simultaneously such that collision dominates in the central regions, and MT dominates in the outer regions \citep{leigh2007blue}. On the other hand, in the case of less dense environments such as OCs and Galactic fields, MT and merger formation channels dominate over collisions \citep{mathieu2015blue}, suggesting the importance of binary systems. Therefore, the fraction of binary formation is unquestionably one of the essential ingredients required to understand the BSS formation process. \cite{sollima2007fraction,milone2012acs, milone2016binary} photometrically calculated the fraction of binary systems in GCs. Depending on the cluster, the estimated fractions of binary systems range from 10$\%$ to 50$\%$. Interestingly, they found that this fraction is strongly correlated with the BSS frequency. On the other hand, these two quantities do not correlate with each other in the case of OCs, unless the groups of clusters with different densities are taken for consideration \citep{cordoni2023photometric}. However, the identification of field blue straggler stars (FBSS) is non-trivial, unlike identifying BSS in star clusters. \citet{preston1994space} identified the field metal-poor stars with MS gravities that are bluer than the MS turn-off of the GCs of comparable metallicity and hence analogous to the BSS found in clusters. Upon comparison, they found that the specific frequencies of 175 blue metal-poor (BMP) stars within $\sim$2 kpc of the Sun are $\sim$9 times greater than those of the BSS found in most metal-poor GCs. This observation implied that cluster-type BSS are just minor components of the field BMPs, and a significant fraction of BMPs are probably the intermediate-age populations of dwarf galaxies that the Milky Way has accreted in the last 10 Gyr. In the subsequent year, however, \cite{preston2000these} discovered that 60$\%$ of the 62 BMP stars that they studied using echelle spectra from the Las Campanas Observatory are binaries. This fraction is roughly four times higher than the normal binaries of the disk and the halo populations \citep{duquennoy1991multiplicity, latham1988survey, latham2002survey}. Furthermore, the mass functions of BMP stars were found to be systematically smaller by a factor of two than those of high proper-motions and Galactic disk binaries. From the above two observations, along with the large periods and low orbital eccentricities of these BMPs, \cite{preston2000these} concluded that a major fraction of BMP stars is FBSS formed via MT. 

Another significant confirmation of the argument that BMP stars are mainly formed via MT was provided by \citet{sneden2003binary}. Using high-resolution spectroscopy, they found BMP binaries showing significant enhancements in carbon and s-process elements such as Sr and Ba, supporting the formation of the BMP binaries by an AGB companion. These were classified as true FBSS. On the other hand, BMP stars having normal/low Sr and Ba abundances than solar were called intermediate-age MS stars. The latter group included both binaries and RV constant stars with 0.25 fraction of binaries, consistent with the disk and halo MS binary frequency \citep{duquennoy1991multiplicity,latham2002survey}. Hence, \citet{sneden2003binary} inferred that RV constant stars are intermediate-age stars, probably accreted from metal-poor dwarf galaxies. 

It is a widely recognised fact that FBSS have a large diversity in the surface chemistry, unlike BSS found in star clusters \citep{andrievsky1996chemical,carney2005metal}. As a result, the binary evolution in the isolation environment of the Galactic fields, as well as the progenitor properties, are very likely to play an important role in the details of MT as well as the chemistry of the accreted material. Therefore, in order to gain a clear understanding of the origin of BMP stars and to confirm whether they are genuine FBSS formed via MT, the detection and characterisation of their hot companions are crucial. The Ultraviolet Imaging Telescope (UVIT) onboard \textit{AstroSat} data has played an incredible role in the discovery of hot companions to BSS in OCs and GCs: post-AGB/horizontal branch star in NGC\,188 \citep{subramaniam2016hot}, white dwarf (WD) in M67 \citep{sindhu2019uvit}, WD in NGC\,5466 \citep{sahu2019detection}, an extreme horizontal branch star (EHB) in NGC\,1851 \citep{singh2020peculiarities}, EHB/subdwarf stars in King\,2 \citep{jadhav2021uocs}, extremely low mass (ELM) WDs in NGC\,7789 \citep{vaidya2022uocs}, ELM and low mass (LM) WDs in NGC\,2506 \citep{panthi2022uocs}, and ELM WDs in NGC\,362 \citep{dattatrey2023globules}. Moreover, normal-mass and high-mass WDs were discovered as companions of yellow straggler stars (YSS) and red clump stars in NGC\,2506 \citep{panthi2022uocs}, and A-type subdwarfs (sdA) as hot companions to YSS in OC NGC\,2818 \citep{rani2023uocs}. This paper focuses on the identification and characterisation of hot companions of FBSS in order to estimate the progenitor properties and their evolutionary stages. 

This paper is organized as follows. The information on observations and data reduction are presented in \S \ref{Section 2}. In \S \ref{Section 3}, the data analysis and results are given. A comprehensive discussion on the possible formation channels of all the BMP stars is presented in \S \ref{Section 4}. The summary and conclusions are given in \S \ref{Section 5}. 

\section{Observation and data reduction} \label{Section 2}
The data used in this work are obtained using UVIT, one of the payloads onboard India's first multi-wavelength space observatory, \textit{AstroSat}, along with other archival data. UVIT has two 38-cm telescopes, one working in the far-UV (FUV) channel (130$-$180 nm), and the other working in both near-UV (NUV) channel (200$-$300 nm), as well as the visible (VIS) channel (350$-$550 nm). The FUV and NUV detectors work in photon counting mode, whereas the visible detector works in integration mode. For further details of the instrument and calibration results, readers are referred to \cite{kumar2012ultraviolet}, \cite{subramaniam2016orbit}, and \cite{tandon2017orbit}. 

\begin{figure*} 
\includegraphics[scale=0.65]{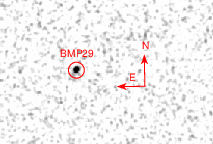}
\includegraphics[scale=0.65]{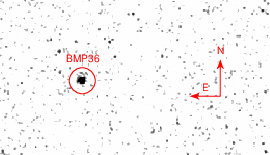}
\includegraphics[scale=0.65]{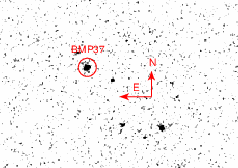}
\includegraphics[scale=0.65]{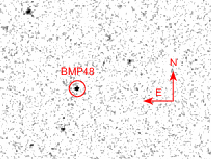} 
\caption{UVIT images of four representative BMP stars, BMP29, BMP36, BMP37, and BMP48 observed in F148W filter.}
\label{Fig.1}
\end{figure*}

We observed 27 BMP stars in two FUV filters, F148W and F169M, using UVIT as part of proposal A10\_053. The observations were carried out between February to September 2021, with final exposure times varying from $\sim$200 sec to $\sim$2000 sec. We have shown the images of four representative BMP stars observed in F148W filter in Figure \ref{Fig.1}.
We used a customized software package, CCDLAB \citep{postma2017ccdlab,postma2021uvit}, to apply spacecraft drift corrections, geometric distortions corrections, flat field corrections, and astrometric corrections to the level 1 (L1) data. The steps followed in CCDLAB in order to obtain the science-ready images are briefly described in the Appendix \ref{Appendix}. Then, we performed aperture photometry on all the science-ready images following curve-of-growth (CoG) technique in CCDLAB \citep{tandon2020additional}. The detailed explanation of this technique is given in Appendix \ref{Appendix2}. 

In order to verify the photometric magnitudes obtained using CCDLAB, we also performed the aperture photometry using the DAOPHOT package in IRAF \citep{stetson1987daophot}. We note that the maximum difference in magnitudes from both the methods is  $\sim$0.05 in both F148W and F169M filters. In order to check the robustness of the magnitudes determination, we performed artificial star tests using \textit{addstar} task under \textit{daophot} package in IRAF. For this purpose, we selected seven representative BMP stars that cover the FUV magnitude range from $\sim$15 to $\sim$21. The recovered magnitudes were found to be comparable to magnitudes of the added stars and the estimated errors in magnitudes obtained after this test are comparable to that obtained by photometry with CCDLAB.

\section{Data analysis and results} \label{Section 3}

\subsection{Spectral energy distributions}

In the last decade, the development of new observing facilities and extensive surveys covering a wide wavelength range of the electromagnetic spectrum has made it possible to generate spectral energy distribution (SED) from ultraviolet (UV) to infrared (IR) wavelengths. These multi-wavelength SEDs of stellar objects can be used to determine their atmospheric parameters, such as effective temperature (T$_{eff}$), radius (R), and luminosity (L). Furthermore, the excess in UV fluxes can indicate the presence of a hotter companion, provided this effect is not due to some of the other factors, such as chromospheric activities or the presence of hot spots. In order to construct the SEDs of BMP stars, we used a virtual observatory SED analyzer (VOSA, \citealt{bayo2008vosa}). Using VOSA, we obtain the photometric fluxes of sources in FUV and NUV from \textit{GALEX} \citep{martin2005galaxy}, optical from \textit{Gaia} DR3 \citep{2021A&A...649A...1G} and PAN-STARRS \citep{chambers2016pan}, near-infrared (NIR) from Two Micron All-Sky Survey (2MASS, \citealt{cohen2003spectral}), and far-infrared from \textit{Wide-field Infrared Survey Explorer} (WISE, \citealt {wright2010wide}). This tool also performs the extinction corrections in the observed fluxes following \cite{fitzpatrick1999correcting} and \cite{indebetouw2005wavelength} in respective bands. It also allows the user to perform a $\chi^{2}$ minimization test by comparing the observed flux to the synthetic flux and identifying the best-fitting spectrum of the theoretical model. The reduced $\chi_{r}^{2}$ is calculated using the following formula: 

\begin{equation} 
   \chi_{r}^{2} =\frac{1}{N-N_{f}}\sum_{i=1}^{N} \frac{(F_{o,i}-M_{d}F_{m,i})^{2}}{\sigma^{2}_{o,i}}
\end{equation} 

where N is the number of photometric points, N$_{f}$ is the number of fitted parameters for the selected model, F$_{o,i}$ and F$_{m,i}$ are the observed and theoretical fluxes, respectively, $\sigma_{o,i}$ are the observational error in the fluxes, and M$_{d}$ is the scaling factor, which must be multiplied with the model to fit the observations. The scaling factor M$_{d}$ is defined as (R/D)$^{2}$ where R is the object radius and D is the distance to the object from the observer. At times the $\chi_{r}^{2}$ values of the fits are large even when the SED fits are visually good, possibly due to some data points with very small observational flux errors, as also noted by \citep{rebassa2021white}. In order to address this issue, VOSA determines another parameter called visual goodness of fit ($vgf_{b}$) which is a modified $\chi_{r}^{2}$. This is calculated by keeping the observational errors to be at least 10$\%$ of the observed fluxes and is determined using the following formula:

\begin{equation}
  vgf_{b} =\frac{1}{N-N_{f}}\sum_{i=1}^{N} \frac{(F_{o,i}-M_{d}F_{m,i})^{2}}{b^{2}_{i}} \\
\end{equation}

where $\sigma_{o,i}\leq {0.1} F_{o,i}$ implies $b_{i} = 0.1 F_{o,i}$ and $\sigma_{o,i}\geq{0.1 F_{o,i}}$ implies $b_{i} = \sigma_{o,i}$. 

In order to construct the SEDs, we performed the following steps: \\

1) We provided the distance of each BMP star from \textit{Gaia} EDR3 data \citep{bailer2021estimating} and extinction values either from the literature or Galactic dust reddening and extinction maps \footnote{https://irsa.ipac.caltech.edu/applications/DUST/}. The distances and extinction values of all BMP stars are listed in Table \ref{Table1} and Table \ref{Table2}. \\
2) We kept two parameters, T$_{eff}$ and log\,\textit{g}, to be free by choosing their ranges to be 3500$-$50000 K and 3$-$5, respectively. \\
3) We fixed the value of metallicity (\big[Fe/H\big]) for all the sources by giving the nearest possible values from the literature. The values of \big[Fe/H\big] of BMP stars are listed in Table \ref{Table1} and Table \ref{Table2}. \\
4) We initially excluded the fluxes in UV wavelengths from SEDs in order to confirm whether fluxes in optical and infrared wavelengths are fitting with the model fluxes. All the extinction-corrected fluxes of BMP stars in different filters are listed in Table \ref{Table3}.\\
5) The above step allowed us to determine if there is excess flux present in the UV wavelength by examining the residuals of the fit, i.e., the difference between the model and observed fluxes.\\
6) We noticed that 10 BMP stars showed the fractional residuals to be nearly zero in all filters and therefore fitted them with single-component SEDs. \\
7) In the case of 17 BMP stars, we noticed the excess to be $>$ 50$\%$ in UV fluxes and therefore tried to fit the double-component SEDs to them. Out of these 17 stars, 12 were successfully fitted using a python based \textsc{Binary SED Fitting} code by \cite{jadhav2021uocs}, which uses $\chi^{2}$ minimization technique. Moreover, we utilised the binary fit feature in VOSA to fit the SED of the BMP2 star using the Kurucz model. However, in the case of the remaining 4 stars, the models used in the above code fitted the UV data points with their highest available temperatures, and the binary fit in VOSA also did not fit the UV data points with any of the available models. Therefore, we show their single-component SEDs.  \\

\subsubsection{BMP stars fitted with the single component SEDs}
 
\begin{figure*}
\includegraphics[scale=0.28]{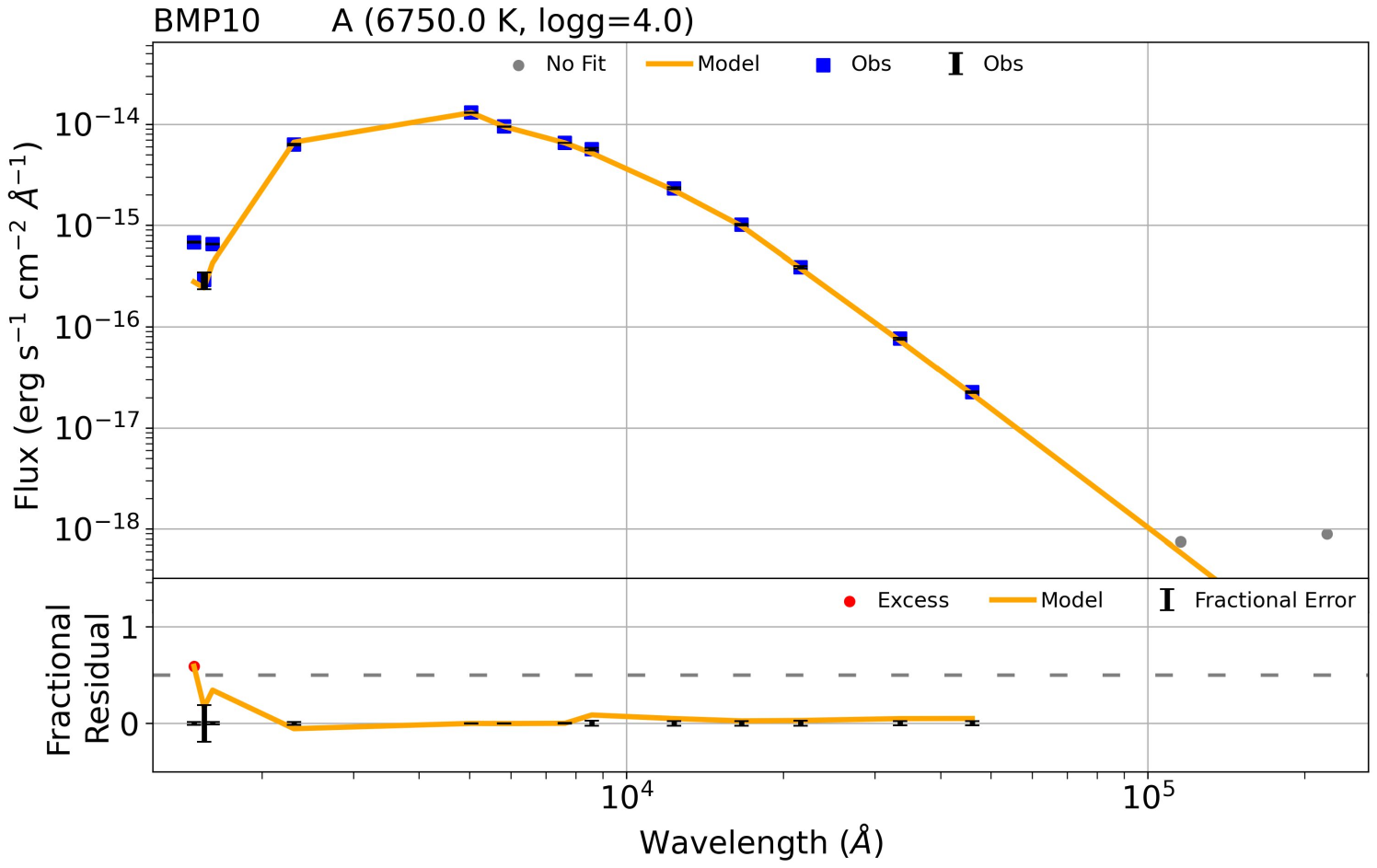} 
\includegraphics[scale=0.28]{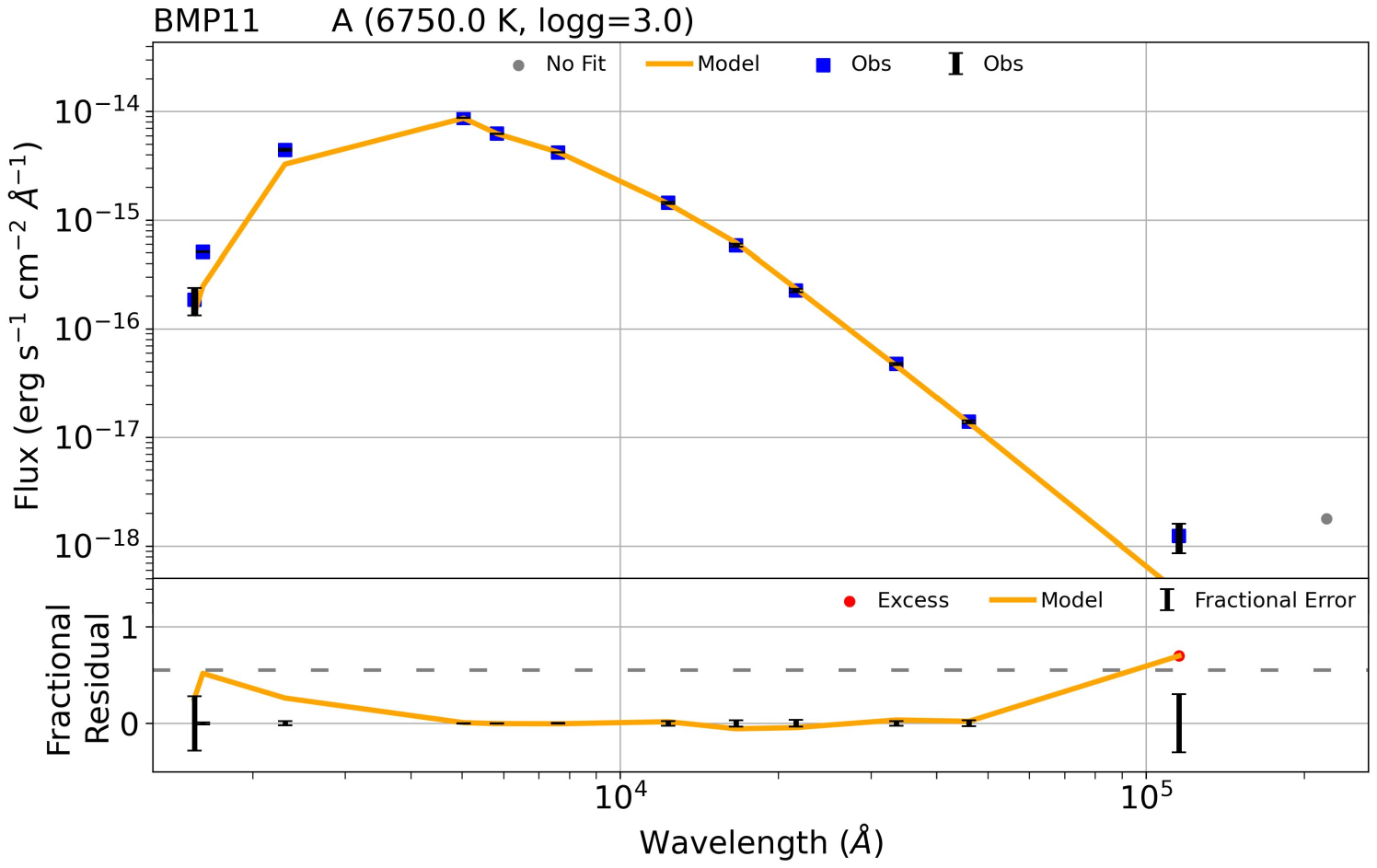}
\caption{The single component SED fits of BMP stars. In the top panel, blue data points show the extinction corrected observed fluxes, labelled as \textit{Obs}. The black error bars represents the errors in observed fluxes and the orange curve represents the Kurucz stellar model fit. The data points which we did not fit either due to upper limits or poor photometric quality are marked as grey-filled circles, labelled as \textit{No Fit}. The bottom panel shows the residual between extinction-corrected observed fluxes and the model fluxes across the filters from UV to IR wavelengths. The excess in the observed fluxes compared to the model fits are shown as red circle, labelled as \textit{Excess}. The orange curve shows the Kurucz stellar model fit and the fractional errors in the residuals are shown in black error bars. The SEDs of other BMP stars fitted with the single component SEDs are shown in Figure \ref{Fig. A1}.}
\label{Fig.2}
\end{figure*}

\begin{figure}
\includegraphics[width=\columnwidth]{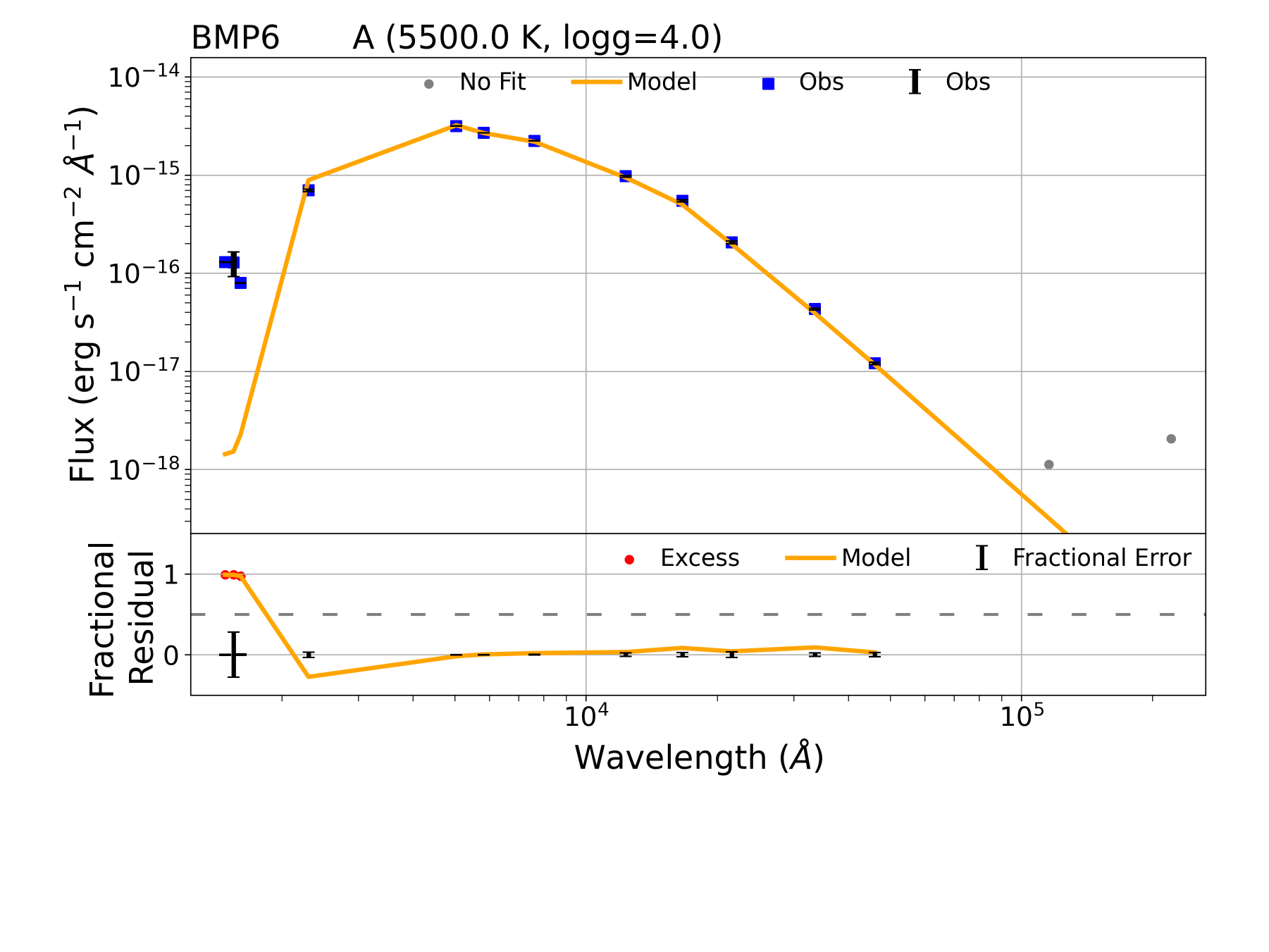}
\caption{The BMP stars showing UV excess but not fitted with the binary-component SEDs. The other SEDs of the BMP stars showing UV excess but not fitted with binary-component are shown in Figure \ref{Fig. A2}. The symbols and curves mean the same as the previous figure.}
\label{Fig.3}
\end{figure}

The single-component SEDs were fitted with Kurucz stellar model \citep{castelli1997notes} and are shown in Fig. \ref{Fig.2}. The top panel for each star shows the fitted SED, whereas the fractional residual in each filter. The values of $vgf_{b}$ parameters in these SEDs are less than 2, which indicates them to be good fits \citep{jimenez2018white,rebassa2021white}. However, as mentioned above, four BMP stars with UV excess $>$ 50$\%$ were not fitted with the binary component SEDs. Therefore, we have shown their single-component SEDs in Fig. \ref{Fig.3}. The parameters of all the BMP stars fitted with the single-component SEDs are listed in Table \ref{Table4}. 

\subsubsection{BMP stars fitted with the double component SEDs}

 We were able to satisfactorily fit the binary component SEDs to 13 BMP stars (Figure \ref{Fig.4}). We utilised the Koester model \citep{koester2010white} to fit the hot component of the SED for 12 stars out of the 13 mentioned earlier. This model provides a temperature range of 5000-80,000 K and a log\,\textit{g} range of 6.5-9.5. For each of these BMP stars, the top panel shows the fitted SED and the bottom panel shows the fractional residual for single and composite fit in all filters. The data points which are not fitted either due to upper limits or poor photometric quality are marked as grey-filled circles. We note that the composite fit compensates for the excess flux in the UV data points, and the residual turns out to be nearly zero in all the data points. This observation is supported by the significant reduction of $\chi^{2}_{r}$ and $vgf_{b}$ values in the case of double fit as listed in Table \ref{Table5}. The errors in the parameters of hot companions were estimated by generating 100 iterations of observed SEDs for each star by adding Gaussian noise proportional to the errors. The median of the parameters derived from the 100 SEDs were considered to be the parameters of hot companions, whereas the standard deviation from the median values were taken as the errors to the parameters. However, in case the statistical error is less than the step size of stellar models, half of the step size is considered as the error. 
 
 We fitted the remaining 1 BMP star, BMP2, using the binary-fit in VOSA, using the Kurucz model as shown in Figure \ref{Fig.5}. Since the temperature of both the companions were known from previous literature \citep{preston1994cs}, we used Kurucz model to fit the double-component SED of this star. In the binary-fit in VOSA, it is assumed that the observed flux is the sum of the fluxes of two different objects, i.e., 
\begin{equation}
F_{obs}(x)\sim M_{d1}*F_{m1}(x) + M_{d2}*F_{m2}(x)
\end{equation}
where F$_{obs}$ is the observed flux, F$_{m1}$ and F$_{m2}$ are the theoretical fluxes from object 1 and object 2, respectively, and M$_{d1}$ and M$_{d2}$ are the scaling factors for object 1 and object 2, respectively. In these SEDs, the data points which are not fitted due to bad photometric quality are shown as yellow circles, and the upper limits are shown as yellow triangles. We note that the observed SED of the BMP2 star was well-fitted using the Kurucz model for both components. The parameters of all the BMP stars fitted with the double-component SEDs are listed in Table \ref{Table5}.

\begin{figure*} 
\includegraphics[scale=0.16]{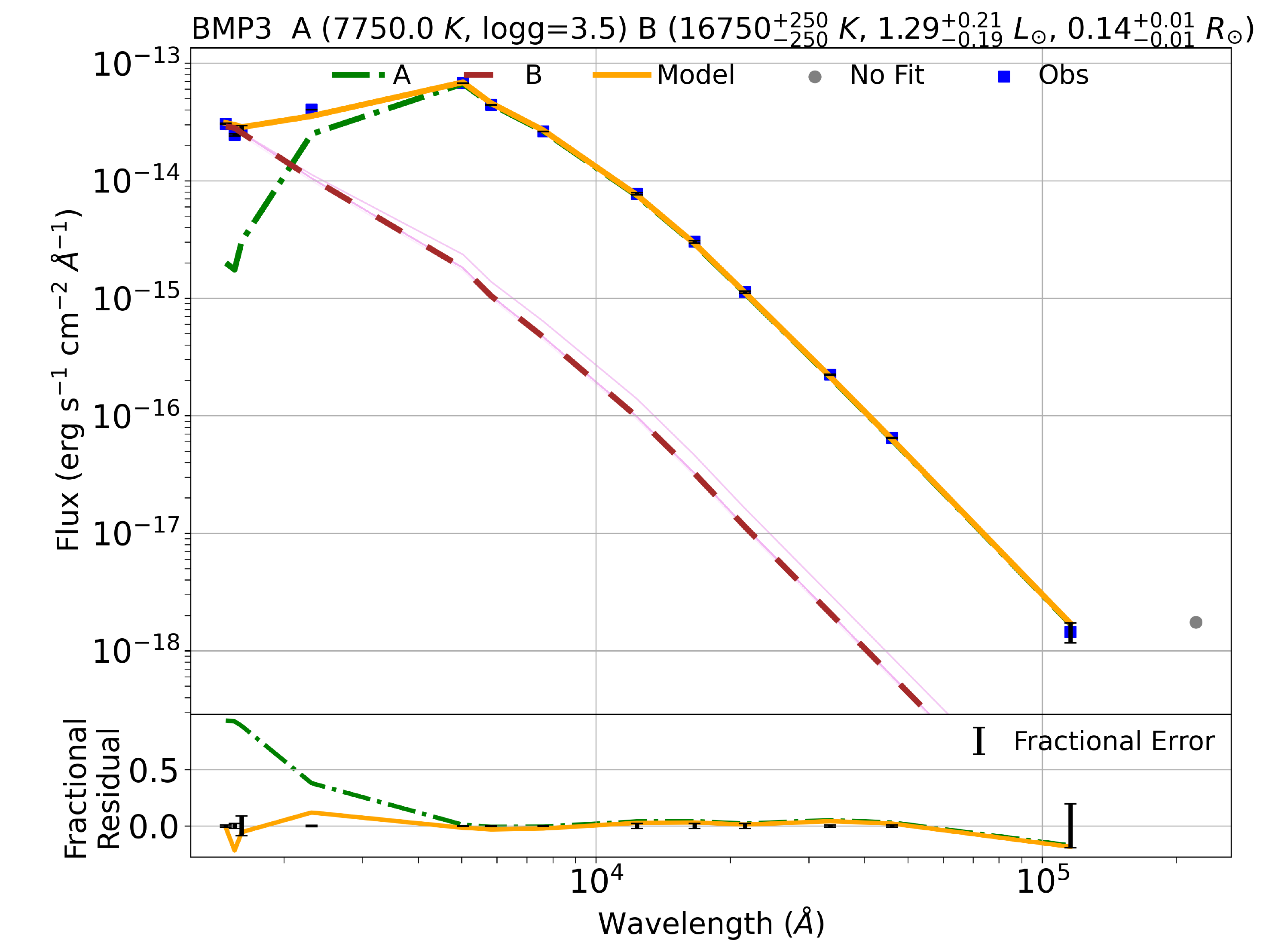} 
\includegraphics[scale=0.16]{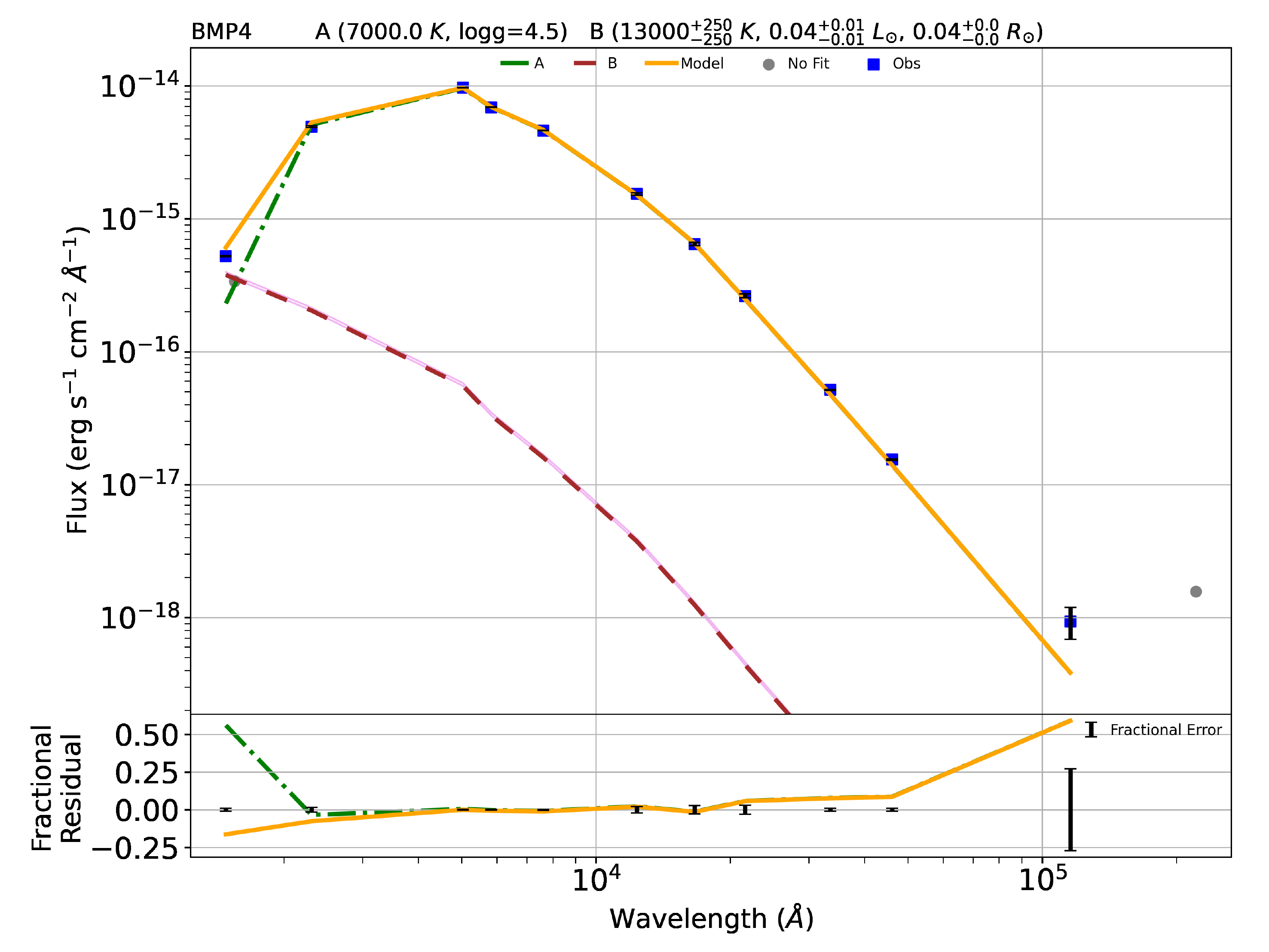} 
\includegraphics[scale=0.16]{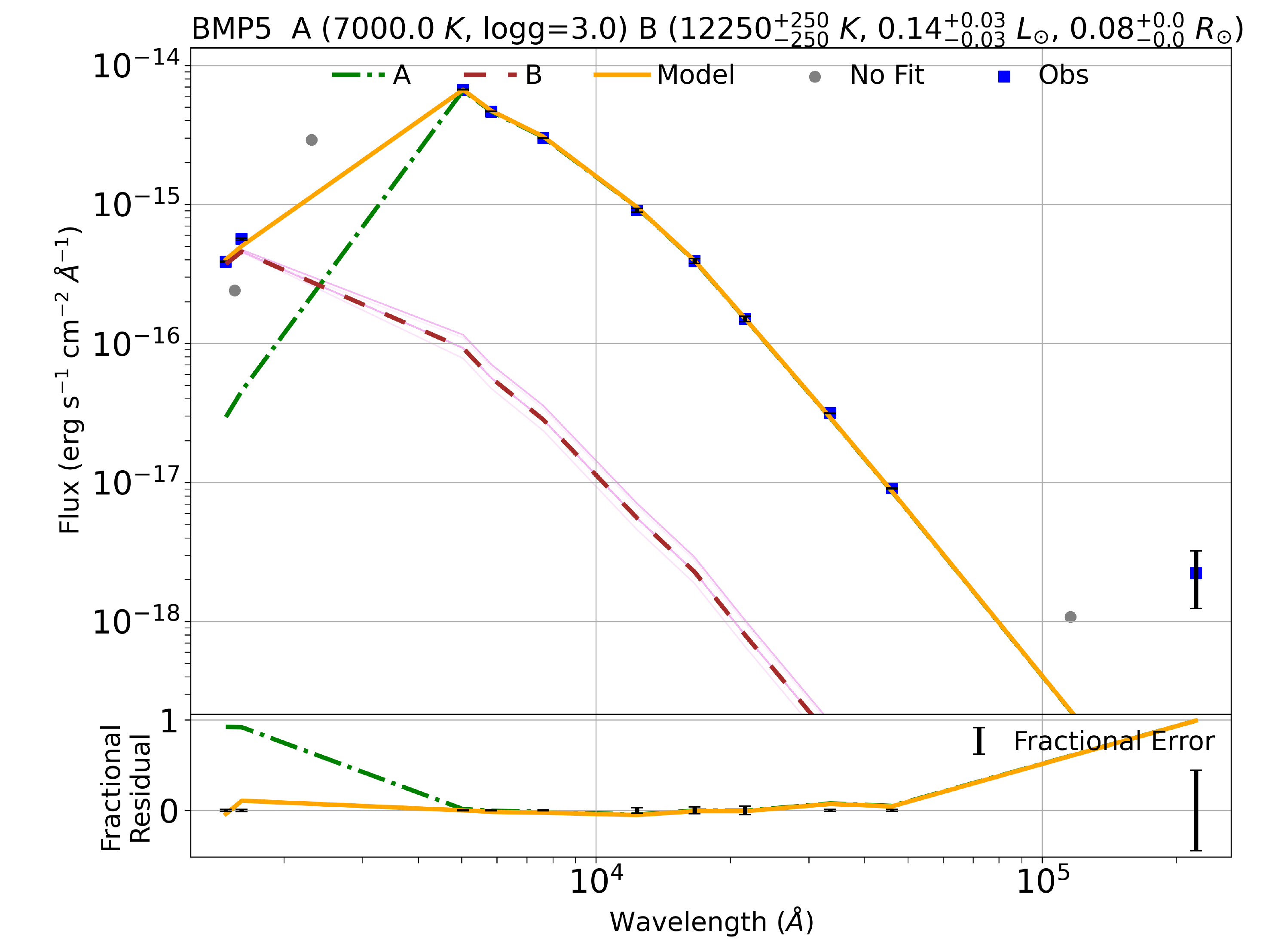}
\includegraphics[scale=0.16]{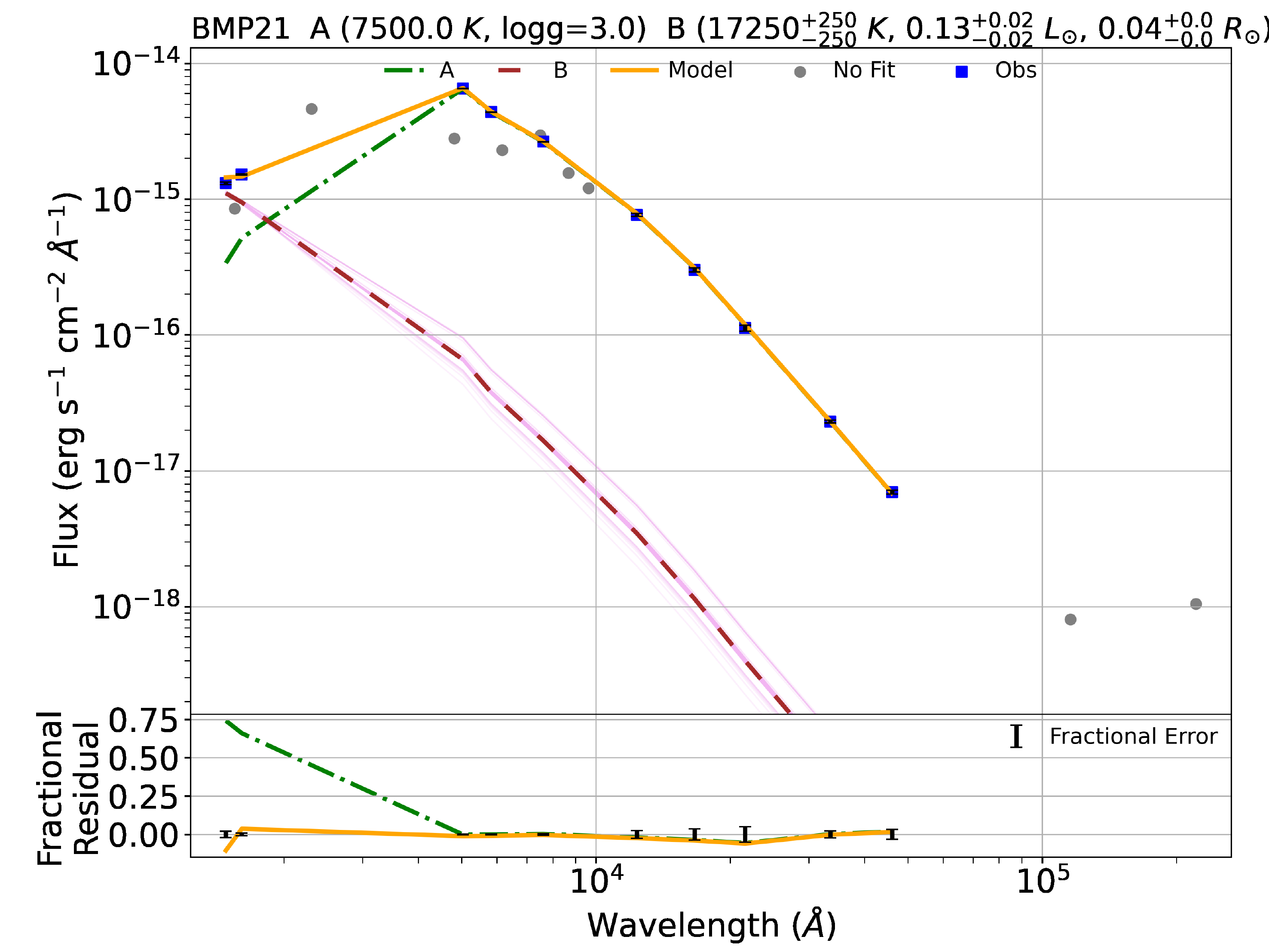} 
\includegraphics[scale=0.16]{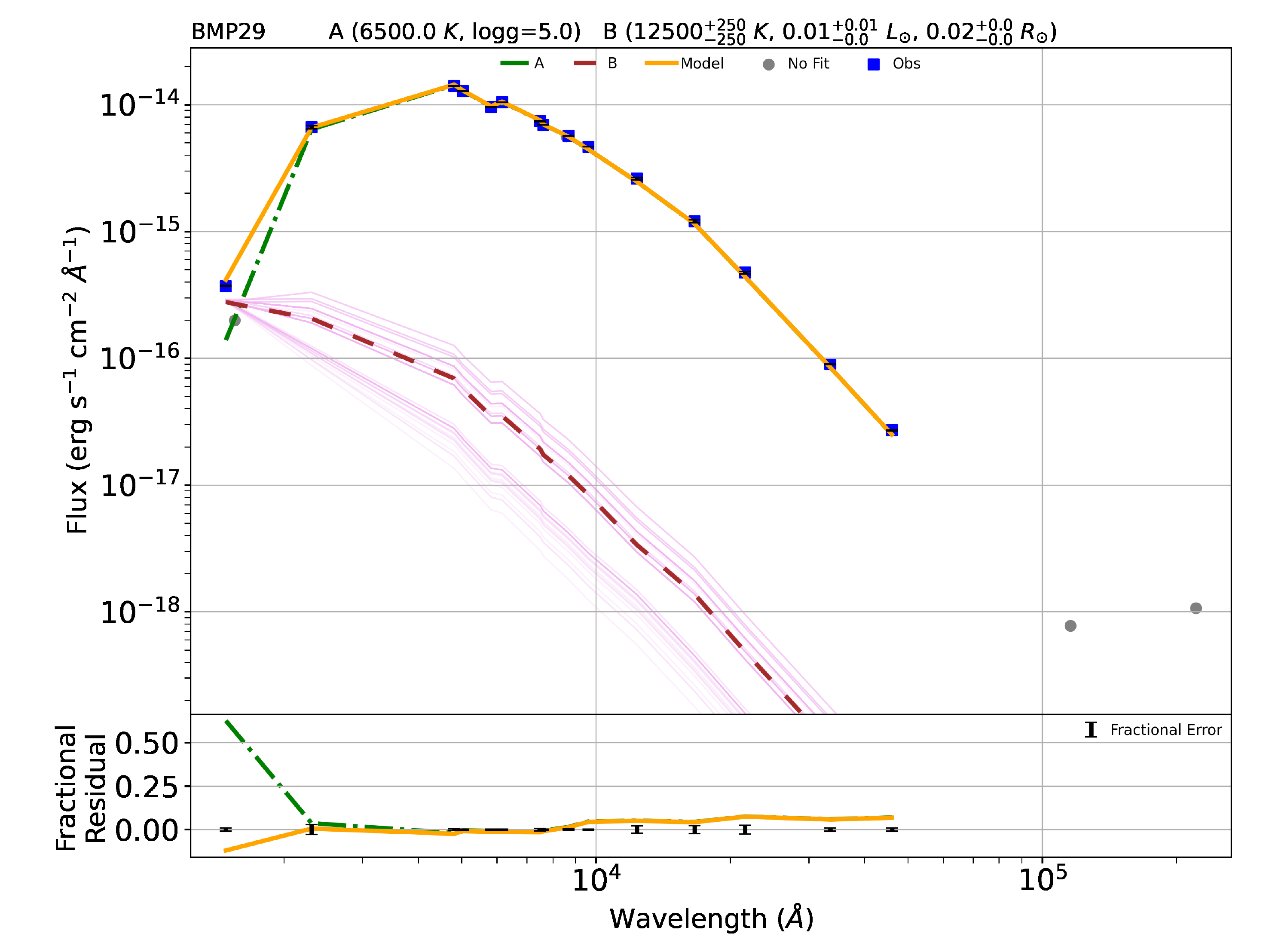} 
\includegraphics[scale=0.16]{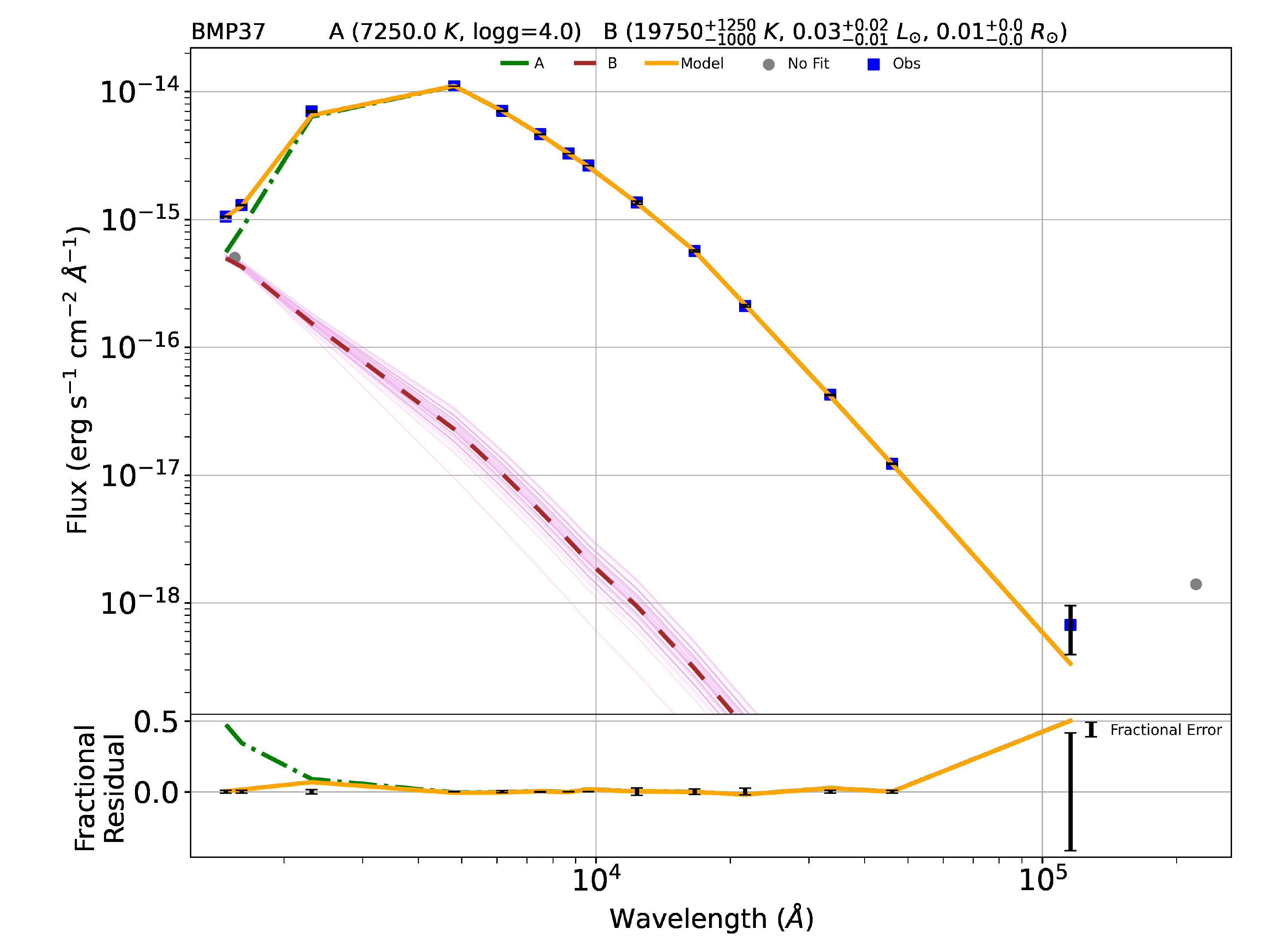} 
\caption{The binary component SEDs of BMP stars. In each of them, the top panel shows the model SED with the cooler (A) component in the green dashed line, the hotter (B) component in the brown dashed line, along with the iterations shown as light pink lines, and the composite fit in the orange solid line. The extinction-corrected flux data points are shown as blue-filled squares with the label \textit{Obs}. The error bars according to flux errors are shown in black bars. The data points not included in the fit are shown as grey-filled circles labelled as \textit{No Fit}. The bottom panel shows the fractional residual for both single fit (green) and composite fit (orange). The fractional errors are shown on the x-axis as black bars with the label \textit{Fractional errors}. The parameters of the cooler and hot companions derived from SED fits, along with their estimated errors, are mentioned at the top of the figures.}
 \label{Fig.4}
\end{figure*}

\begin{figure*}  \ContinuedFloat
\includegraphics[scale=0.16]{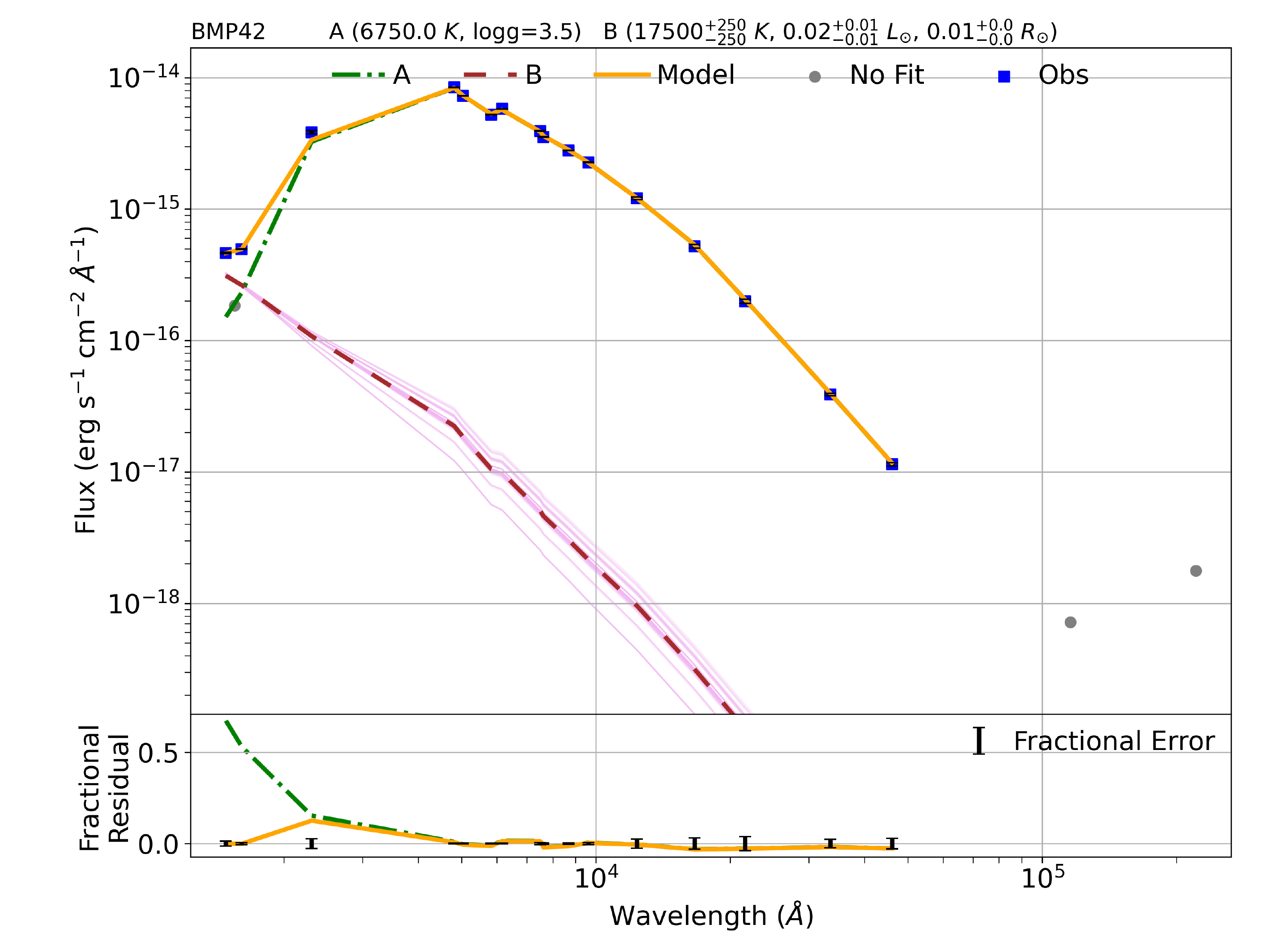}
\includegraphics[scale=0.16]{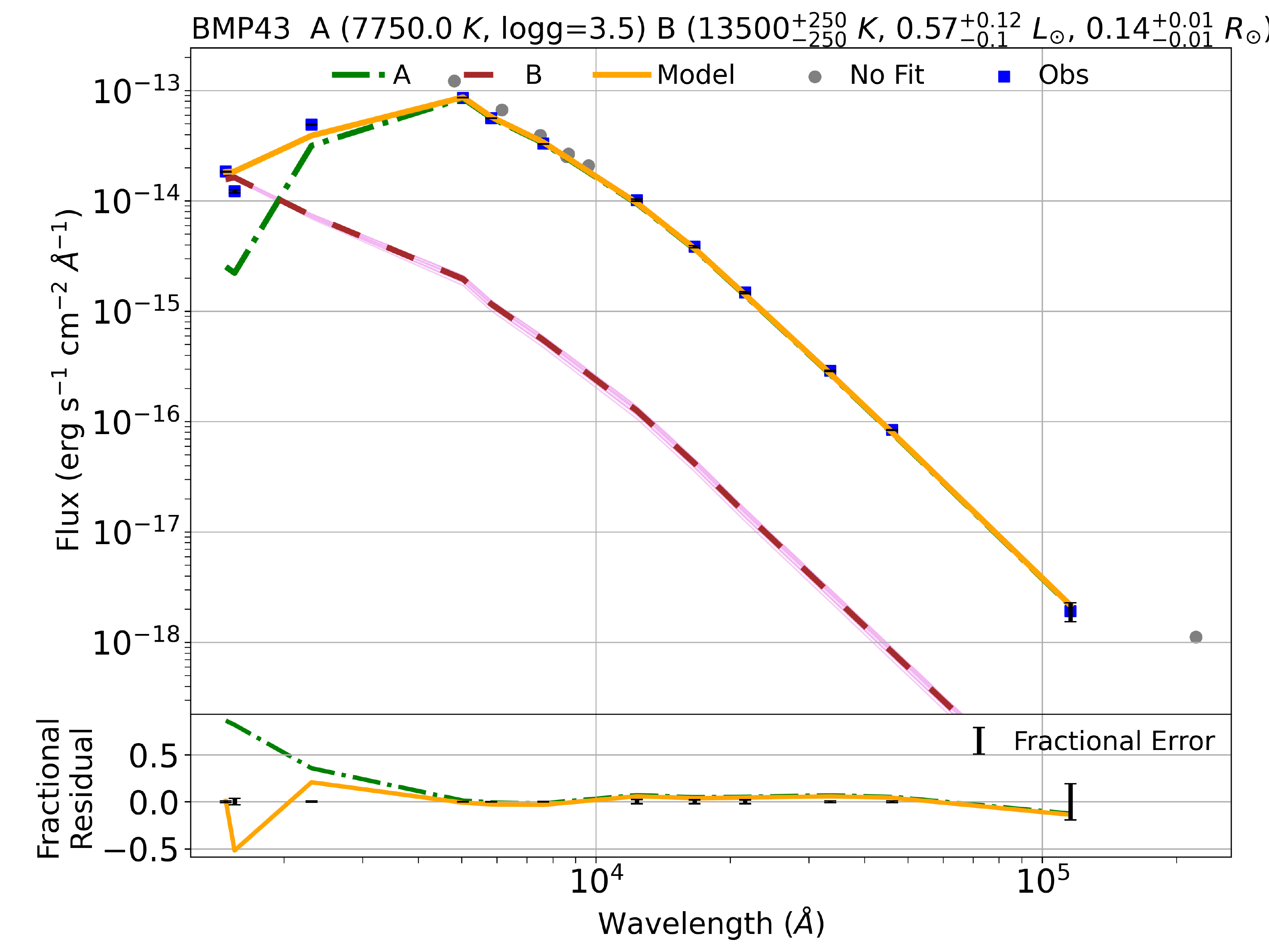}
\includegraphics[scale=0.16]{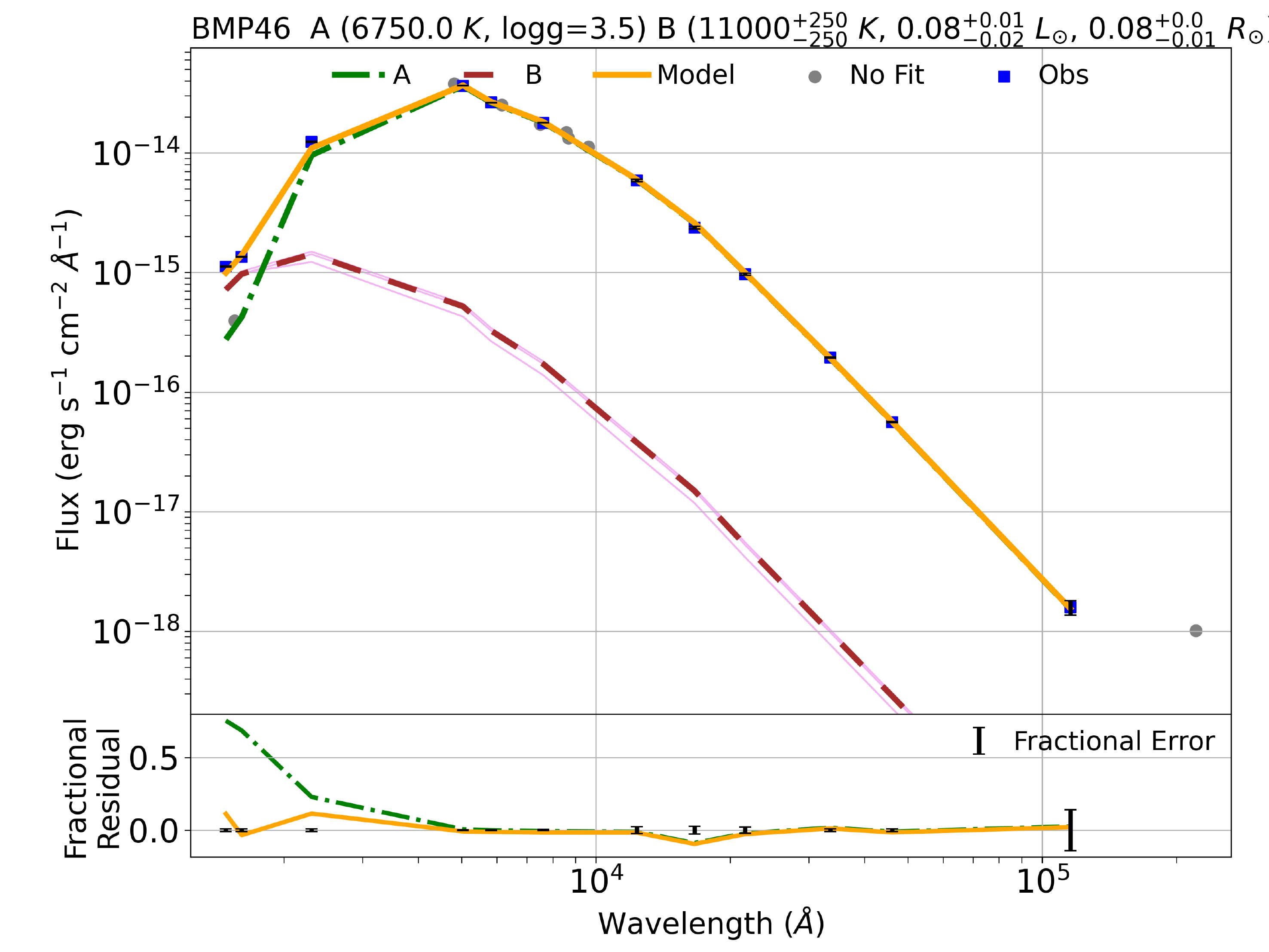}
\includegraphics[scale=0.16]{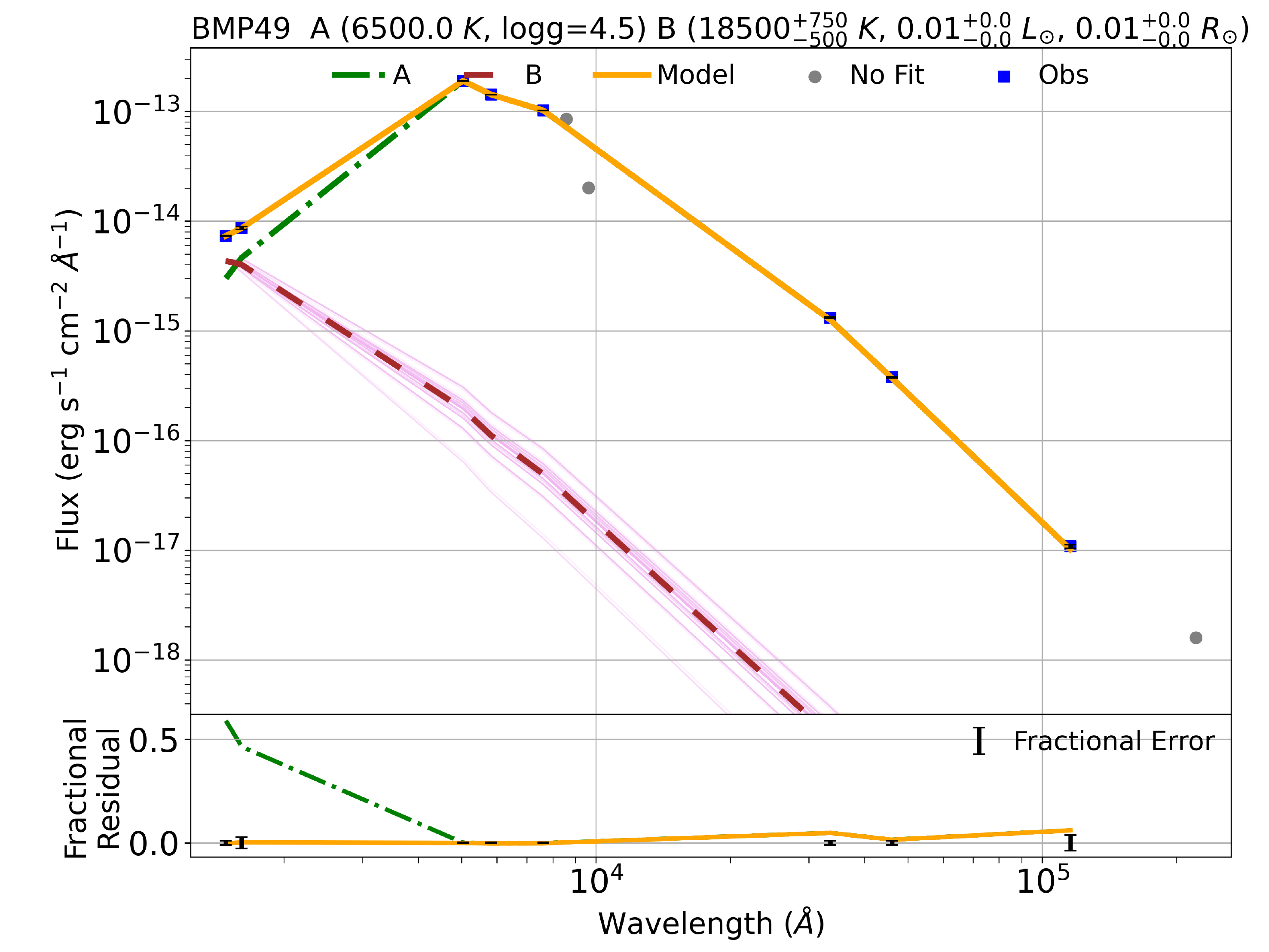} 
\includegraphics[scale=0.16]{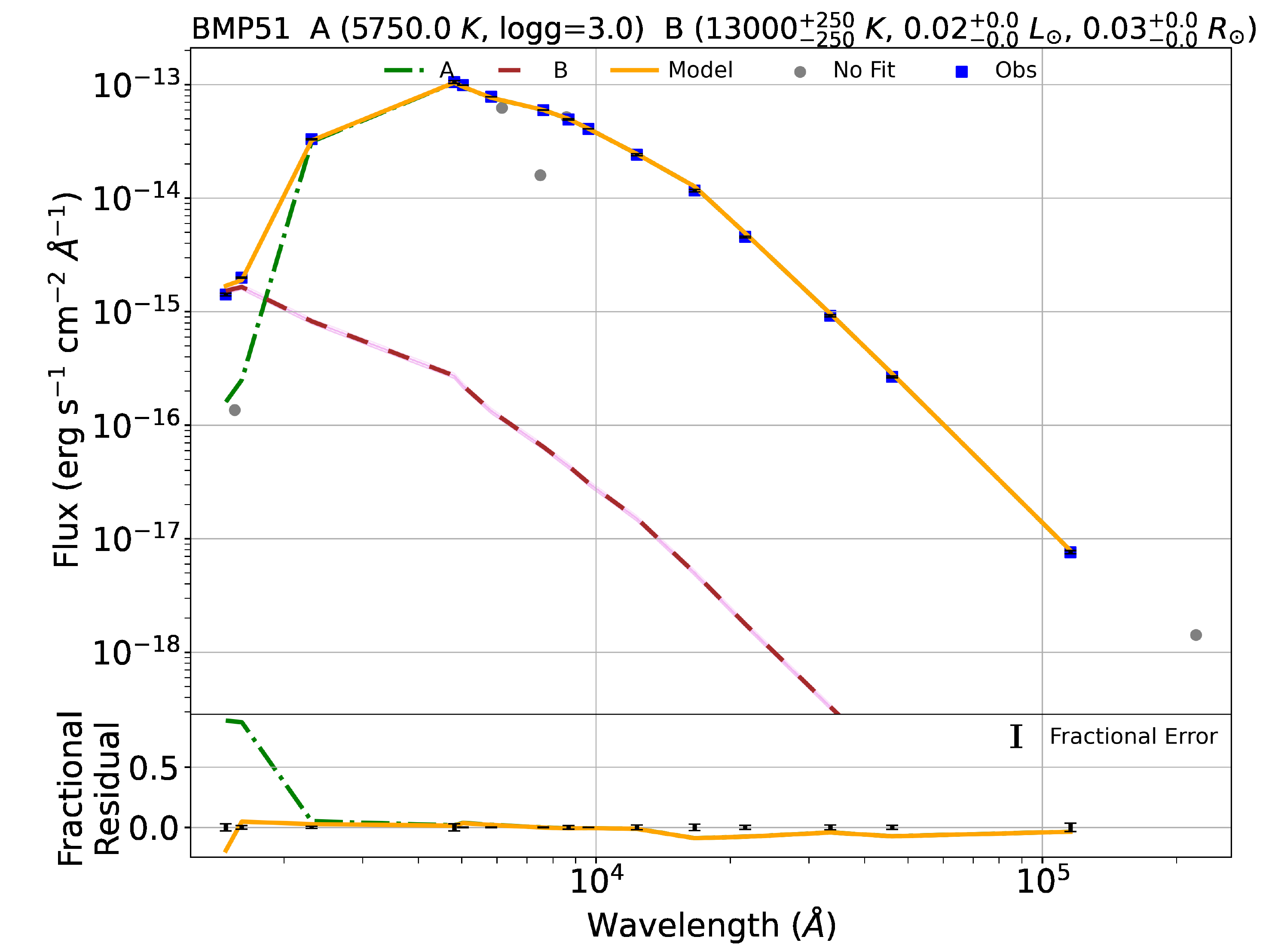} 
\includegraphics[scale=0.16]{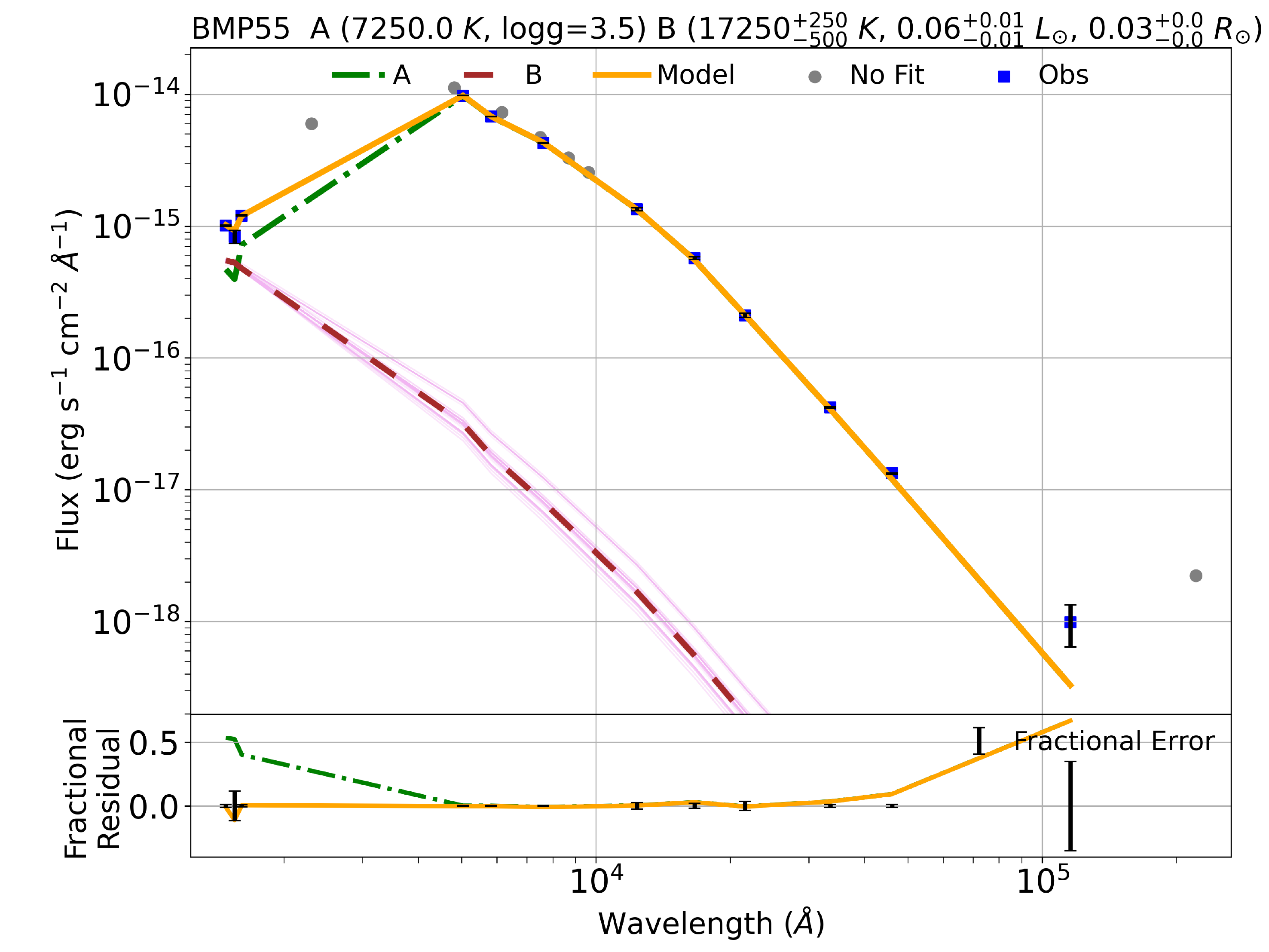} 
\caption{The binary component SEDs of BMP stars (cont.)}
\label{Fig.4}
\end{figure*}

\begin{figure}  
\includegraphics[scale=0.36]{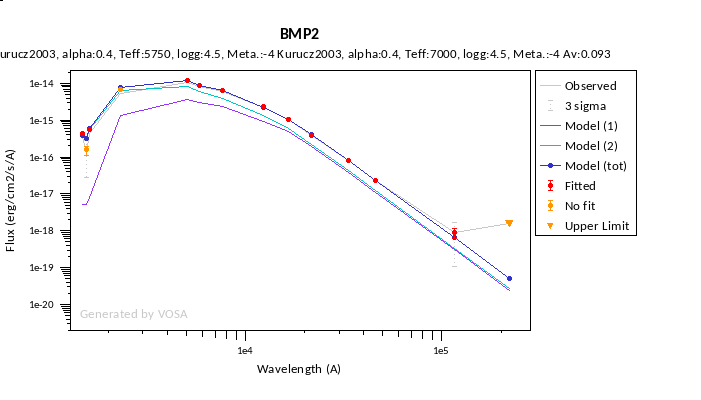}
\caption{The binary component SED of BMP2 fitted using VOSA. The observed fluxes are shown as red-filled circles, model fluxes in blue-filled circles, data points which are not fitted due to bad photometric quality in yellow circles, and the upper limits in yellow triangles. The cooler (A) component fit is shown in the purple line, the hotter (B) component fit in the light blue line, and the composite fit in the dark blue line. At the top of the figure, the models of both the companions, along with their temperatures and log\,{\it g}, are mentioned.}
\label{Fig.5}
\end{figure}

\subsection{Classification of BMP stars as thick disk and halo stars}

We followed the criterion by \citet{bragaglia2005chemical} to separate thick disk and halo stars on a kinematical basis using Galactic space velocity ($\sqrt{U^{2} + V^{2} + W^{2}}$) and \big[Fe/H\big]. \cite{bragaglia2005chemical} categorized sources with space velocities $>$ 100 kms$^{-1}$ and \big[Fe/H\big] $<$ --1 as halo stars, whereas sources with space velocities $<$ 100 kms$^{-1}$ and \big[Fe/H\big] $>$ --1 as thick disk stars. We calculated space velocities of all the BMP stars using ``gal$\_$uvw'' function in PyAstronomy package \citep{czesla2019pya} \footnote{https://pyastronomy.readthedocs.io} for which parallaxes, PMs, and RVs were given as input. This function takes the general outline of \cite{johnson1987calculating} except that U is positive outward toward the Galactic anti-center. We applied the algorithm and transformation matrices of \cite{johnson1987calculating} to obtain the Galactic space velocity components (U,V,W). The parallaxes and PMs are taken from \textit{Gaia} DR3 \citep{babusiaux2022gaia}, whereas RVs are taken from both literature and \textit{Gaia} DR3 data. We used RV values from \textit{Gaia} DR3 for BMP 49 and BMP 50 as they were not available in the literature, whereas for all the other stars, we use literature values. Apart from BMP 49 and BMP 50, we compare the RV values from \textit{Gaia} as well as those available in literature and found that the RV values are within the range of $\pm$ 5km/s.

\begin{figure}
\includegraphics[width=\columnwidth]{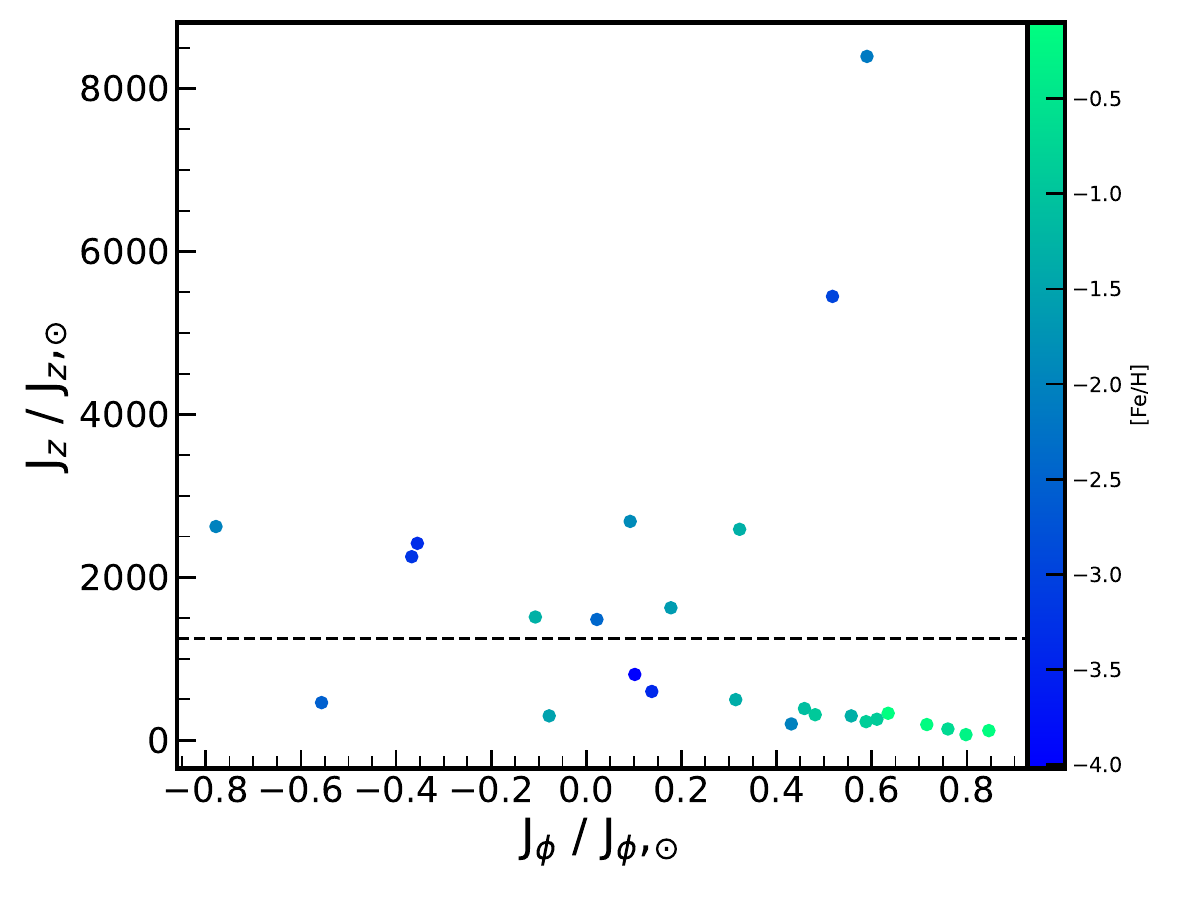}
\caption{The azimuthal action vs vertical action normalized by the solar values with the metallicity in the color bar. The horizontal dashed-dotted line indicates J$_{Z}$ / J$_{Z}$, $_{\phi}$ = 1250.}
\label{Fig.6}
\end{figure}

We also followed \cite{cordoni2021exploring} to verify the above classification of the stars. We use the GALPY \footnote{http://github.com/jobovy/galpy} Python package \citep{bovy2015galpy} to determine the kinematics of our stars. We use the same orbit as used by \cite{cordoni2021exploring}, obtained by integration backward and forward in time for 2 Gyrs. It comprises of an asymmetric model with a bulge, thin and thick disks. We calculate the vertical actions ($J_{z}$) and azimuthal actions ($J_{\phi}$) using the 6D parameters from \textit{Gaia} DR3, and normalized by the solar values, $J_{\phi,\odot}$ = 2014.24 kpc km/s and $J_{z,\odot}$ = 0.302 kpc km/s as used in \cite{cordoni2021exploring}. Actions are considered highly valuable because they are nearly conserved, assuming a smoothly evolving potential \citep{binney1984spectral}. The significance of using actions are described in detail in studies such as \cite{myeong2018milky} and \cite{trick2019galactic}. We plot the J$_{\phi}$ vs J$_{z}$ normalized by solar values for all our 27 BMP stars as shown in Figure \ref{Fig.6}. The dashed-horizontal line marks the region with J$_{z}$/J$_{z,\odot}$ = 1250. All the sources above this line are halo stars, whereas the sources shown below this line are thick disk stars, as reported in \cite{sestito2019tracing} and \cite{cordoni2021exploring}. Except for BMP29, both of the above mentioned methods agree in the classification of the BMP stars as thick-disk and halo stars. BMP29 is a likely halo star using the \big[Fe/H\big] and space velocities criteria, whereas a thick disk star using the analysis based on Galactic actions. However, we consider it a halo star in the rest of the analysis.

The parameters of thick disk stars fitted with the double-component SEDs are tabulated in Table \ref{Table6}, whereas the parameters of halo stars fitted with the double-component SEDs are tabulated in Table \ref{Table7}.

\section{Discussion} \label{Section 4} 

\begin{figure}
\includegraphics[width=\columnwidth]{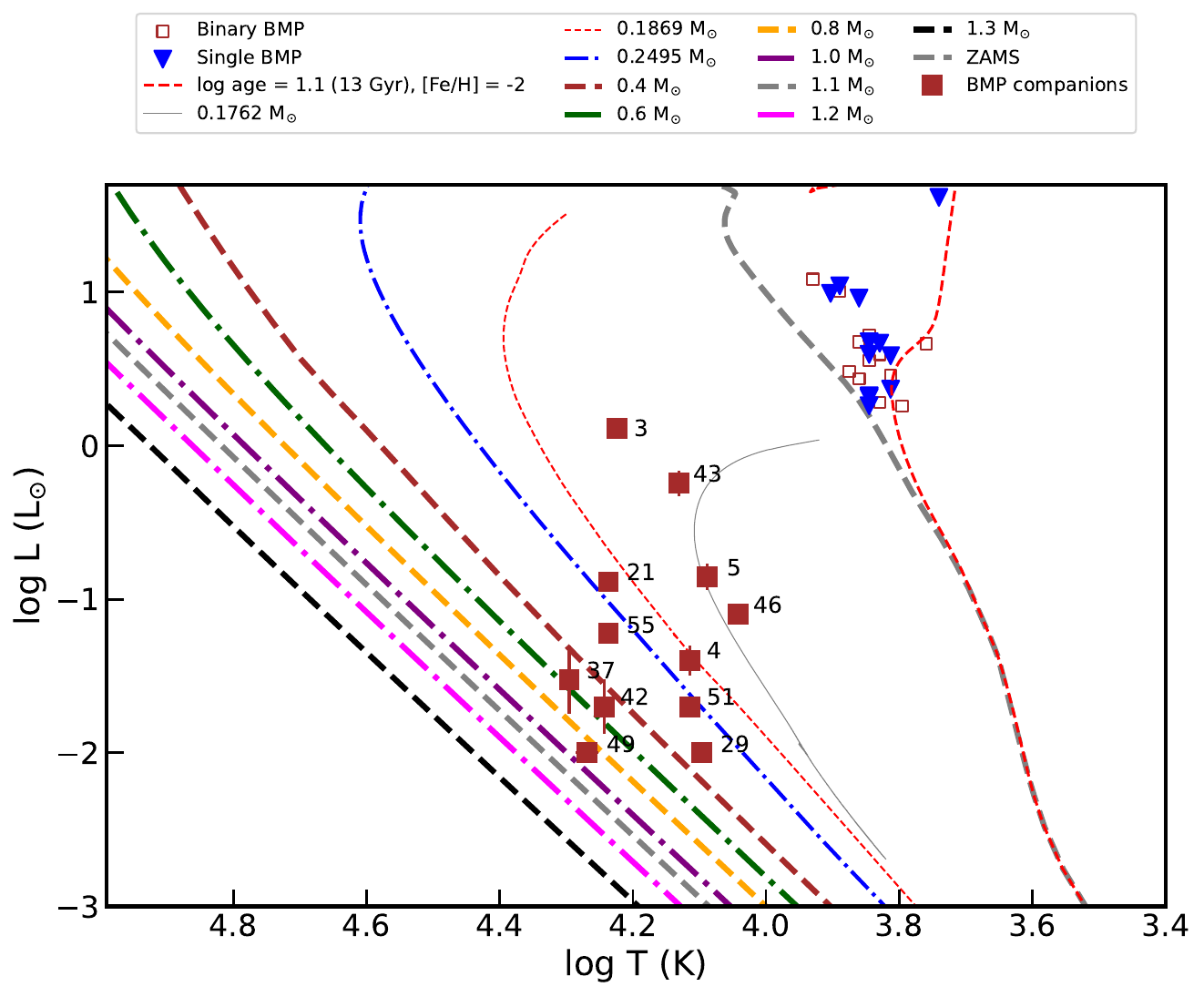}
\includegraphics[width=\columnwidth]{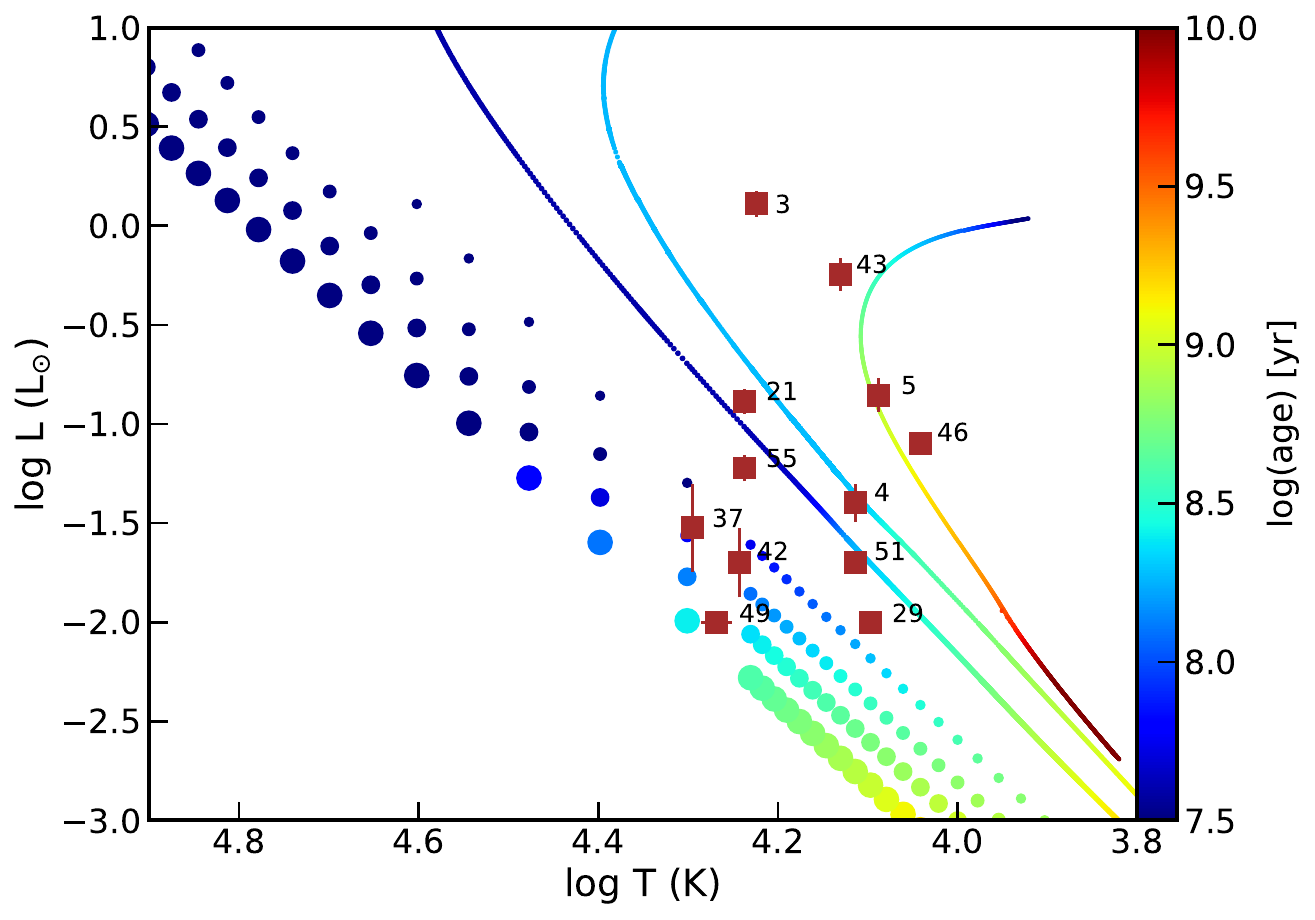}
\caption{The upper panel represents the H-R diagram showing the single component BMP stars as blue  triangles, the cooler companion of BMP stars as brown open squares, and their corresponding hot companions as brown-filled squares. A PARSEC isochrone of 13 Gyr age is plotted as the red dashed curve, and ZAMS is plotted as the grey dashed curve. The WD cooling curves of different masses taken from \citet{panei2007full} and \citet{althaus2013new} are represented by dashed curves of different colours. The lower panel represents the hot companions of BMP stars lying on WD cooling curves indicating their approximate cooling ages as shown in the colour bar on the right.}
\label{Fig.7}
\end{figure}

In order to understand the formation mechanisms of BMP stars, the nature of their hot companions should be investigated. In this context, we plotted the Hertzsprung-Russell (H-R) diagram as shown in Figure \ref{Fig.7}. The upper panel shows a PARSEC isochrone of 13 Gyr age and \big[Fe/H\big] = --2. We have shown all the BMP stars fitted with single as well as binary component SEDs. In the case of binary component SEDs, both cooler and hot companions are shown. Furthermore, a zero-age MS (ZAMS) and WD cooling curves \citep{panei2007full, althaus2013new} of extremely low-mass (M $<$ 0.2 M$_{\odot}$), low-mass (M $\sim$0.2 -- 0.4 M$_{\odot}$), normal-mass (M $\sim$0.4 -- 0.6 M$_{\odot}$), and high-mass (M $>$ 0.6 M$_{\odot}$) are also shown.

It is noteworthy that the hot companions of BMP stars show a large spread in their masses varying from M $\sim $0.17 -- 0.8 M$_{\odot}$. This discovery, when combined with the known chemical composition of individual BMP stars, can unveil the evolutionary pathways which their progenitors may have gone through. In this section, we discuss the known properties and chemical compositions of the BMP stars, as mentioned in the literature (Table \ref{Table1} and Table \ref{Table2}), along with the implications of fitting them with single or double component SEDs. The enhancements and deficiencies in all the elements mentioned hereafter refer to their abundances greater or lesser than the solar values, respectively. Furthermore, we discuss the correlation of thick disk/halo stars with the nature of the corresponding hot companions.

\subsection{Properties of BMP stars fitted with the single component SEDs}
Among the BMP stars fitted with the single component SEDs, BMP10 and BMP20 are RV constant stars and slow rotators (\textit{vsini} $<$ 25 km/s). BMP10 is deficient in \big[Sr/Fe\big], whereas BMP20 is deficient in \big[Ba/Fe\big] \citep{preston2000these}. These may imply that they are probably intermediate-age MS stars accreted from dwarf satellite galaxies \citep{sneden2003binary}.

Other stars fitted with the single-component SEDs include BMP11, BMP17, BMP23, and BMP30, which are SB1s' with periods and eccentricities as listed in Table~\ref{Table1}. Moreover, BMP11, BMP17, BMP23 are enhanced in \big[Sr/Fe\big] and BMP23 is enhanced in \big[Ba/Fe\big] \citep{preston2000these}. Therefore, these are candidates for FBSS \citep{sneden2003binary}. These stars may harbour relatively fainter/cooler WDs, not detected by UVIT. In the case of BMP11, we noticed an excess in WISE/WISE.W3 and WISE/WISE.W4 fluxes. We found a nearby source present within 3$\arcsec$ of this star in \textit{Aladin}\footnote{https://aladin.u-strasbg.fr/} in allWISE filter, likely to be responsible for the observed excess in IR data points. Another star, BMP15, is a known SB1 but deficient in \big[Sr/Fe\big] and \big[Ba/Fe\big], likely intermediate-age MS star \citep{sneden2003binary}. Two stars, BMP44 and BMP50, are also known to be SB1s and rapid rotators \citep{preston2000these,carney2005metal}. However, no information about their abundance of Sr and Ba is available in the literature. The absence of excess UV fluxes in all the above-known SB1 systems implies that their probable WD binary companions are not bright enough to be detected with UVIT. \\

BMP36 has an extreme enhancement of lead (\big[Pb/Fe\big] = +3.7), the highest seen in any star so far. Such overabundances are the signatures of AGB and post-AGB evolution of the companion stars, now likely to be WDs. It is a rapid rotator and abundant in \big[Sr/Fe\big] and \big[Ba/Fe\big] \citep{preston2000these}. \cite{bonifacio2009first} observed this star with the ESO VLT and the high-resolution spectrograph UVES at a resolution of R = 45,000 and noted that this is an SB2 star. We fitted a single component SED to this star, indicating the non-detection of a hot companion. This also suggests that if it is an SB2 system, the T$_{eff}$ of the two components are unlikely to be very different. If this harbours a cooler/fainter WD not detected by the UVIT, then this is likely to be a triple system.  \\

BMP6, BMP13, and BMP48 are SB1s and deficient in \big[Sr/Ba\big] and \big[Ba/Fe\big] elements \citep{preston2000these, bonifacio2009first}. However, BMP14, which is also a known SB1, is deficient in \big[Sr/Fe\big] but enhanced in \big[Ba/Fe\big]. These stars are likely to be intermediate-age MS stars as per the criteria mentioned by \cite{sneden2003binary}. If they harbour WDs, they probably are below the detection limit of UVIT. BMP6 is a carbon-enhanced metal-poor (CEMP)-no object star, i.e., a subclass of CEMP stars that have no strong over-abundances of s-process elements \citep{yong2012most} and therefore cannot easily be explained by Case-C MT from a binary AGB companion. This combination of the metal-poor stars being enhanced in carbon and not in s-process elements suggests that their abundance may be intrinsic \citep{meynet2006early, meynet2010c, chiappini2013first}. However, it is also possible that some CEMP-no stars have been polluted by a companion, but the binary fraction of CEMP-no stars is not well constrained so far. Moreover, the BMP6 star is classified as a binary candidate by \citet{arentsen2019binarity} due to variations in RV with a period of 300 days.
BMP48 is among the extremely metal-poor (\big[Fe/H\big] $<$ $-$2.5) stars \citep{bonifacio2009first}, which are important in understanding the early chemical evolution of the Galaxy. While fitting the SED of this source, we note that UV fluxes show $>$ 50$\%$ excess, indicating the presence of a hotter companion. However, we are not able to obtain a reliable fit to the hotter part of the SED. 

In summary, some of the BMP stars listed here could still have a remnant of an evolved star not hot enough to detect with UVIT.  

\subsection{Properties and formation mechanisms of BMP stars fitted with the double component SEDs}

\textit{BMP2 (BPS CS 22873-0139)}: \citet{preston1994cs} carried out the first detailed study of this highly metal-deficient (\big[Fe/H\big] = $-$3.4) star based on low-resolution (R $\sim$16,000) spectroscopy. They reported it to be an SB2 with a period of 19.16 days, an eccentricity of 0.20, and a mass ratio of $\sim$0.88. Following the definition by \citet{preston1994space}, they classified the primary component as a BMP star, with B--V = 0.37\,mag and U--B = $-$ 0.22\,mag. Moreover, they compared the estimated luminosity ratios of the two components with the measured mass ratio and suggested the age of the system to be $\sim$8 Gyr. Since such a young age is unexpected for very metal-poor stars, they concluded that BMP2 must have accreted from a low-luminosity satellite of the Milky Way. Later, \citet{spite2000element} performed the high-resolution (R $\sim$50,000) spectroscopy in order to determine its element abundances and noted a very low abundance of Sr and $\alpha$-elements such as Mg and Ca. Moreover, they estimated the temperatures of both the binary components to be 5750 K and 6300 K, suggesting them to be MS stars. They reported that the most plausible explanation for the formation of this star could be that its progenitor(s) is (are) one (several) hypernova(e). Furthermore, \cite{sneden2003binary} classified this star as an intermediate-age MS star based on the deficiency in \big[Sr/Fe\big] and \big[Ba/Fe\big] abundances determined using high-resolution echelle spectra.  

As mentioned above, we used the Kurucz model in order to fit both the components by giving a temperature range of 5000$-$5750 K and 6000$-$7000 K using binary-fit in VOSA. We note that the model fits satisfactorily for both the components as shown in Figure \ref{Fig.4}. The T$_{eff}$ estimated from the SEDs are in agreement with the previous estimations. Hence, we confirm that BMP2 is an SB2 system, where both companions are MS stars.\\  

\textit{BMP3 (TYC 8778-1253-1)}: \citet{preston2000these} classified this star as an SB1 with a period of 1.23 days, circular orbit and a rapid rotation of \textit{vsini} = 60 km/s. This star shows enhancements in \big[Sr/Fe\big] and \big[Ba/Fe\big], and therefore classified as FBSS by \cite{sneden2003binary}. From the parameters derived using the best-fit SED, we note that the hotter companion of this star is an ELM WD of mass $\sim$0.16M$_{\odot}$, as shown in the upper panel of Figure \ref{Fig.7}. This ELM WD is likely to have formed via case-A/case-B mass transfer since neither ELM nor LM WDs can form via single stellar evolution within Hubble time \citep{brown2011binary}. Hence, we confirm that BMP3 is indeed an FBSS. We also note that this star is likely to belong to the thick disk. However, this MT is unlikely to produce the observed chemical enhancements in \big[Sr/Fe\big] and \big[Ba/Fe\big] since they are the hallmarks of the contamination via the AGB companion (Case-C MT).\\

\textit{BMP4 (BPS CS 22896-0173)}: \citet{wilhelm1999spectroscopy} performed  medium-resolution spectroscopy and UBV photometry for a sample of 1121 A-type stars in the halo and disk. There were 115 stars in their sample that were previously classified as BMPs by \cite{preston1994space}, most of which fell into the MS A-type category, including BMP4. Although MS gravity of A-type stars have lifetimes that are significantly shorter than the expected turnoff age of the halo population, it has been suggested that these stars exist in the metal-poor thick disk \citep{lance1988young}, and even as members of the metal-poor halo \citep{preston1994space}. In addition, BMP4 is known to be a slow rotator (\textit{vsini} = 12 km/s) and an RV constant star \citep{preston2000these}. The parameters of the hotter companion derived from the best-fit SED suggest that the hotter companion of BMP4 is an ELM WD of mass $\sim$0.18 M$_{\odot}$ (upper panel of Figure \ref{Fig.7}), implying that BMP4 is an FBSS formed via Case-A/Case-B MT. This result is consistent with the observed deficiency in \big[Sr/Fe\big] and \big[Ba/Fe\big] as reported by \cite{preston2000these}. We also note that this star is likely to belong to the thick disk. However, it may be possible that the inclination of the binary orbits is such that the variations in RV were not observed. The cooling age of the WD companion of this star is $\sim$0.31 Gyr, and the star does not show rapid rotation.  \\

\textit{BMP5 (BPS CS 22896-0103)}: This star is a binary candidate with a period of 31.66 days and an eccentricity of 0.45. It is a rapid rotator (\textit{vsini} = 45 km/s) and deficient in \big[Sr/Fe\big] and \big[Ba/Fe\big] \citep{preston2000these}. The parameters of the hotter companion estimated by constructing the SED suggest it to be an ELM WD of mass  $\sim$0.16 M$_{\odot}$ (upper panel of Figure \ref{Fig.7}). In addition, the WD cooling curves suggest that the WD companion of this star formed $\sim$1 Gyr ago. An ELM WD in a short-period binary makes this star an FBSS that has formed via Case-A/Case-B MT. This conclusion is in agreement with the binary nature, along with the fact that this star is deficient in Sr and Ba. This is likely to be an intermediate-age system belonging to the thick disk. We noticed a relatively high eccentricity for this post-MT system, as this may be in a triple system or belong to the class of binaries that are ejected from star clusters \citep{khurana2022dynamically}.\\

\textit{BMP21 (BPS CS 29518-0024)}: This is an SB1 with a period of 64.7 days and eccentricity of 0.44, a rapid rotator (\textit{vsini} = 40 km/s), and enhanced in \big[Sr/Fe\big] and \big[Ba/Fe\big] \citep{preston2000these}. From the best-fit parameters derived by constructing the SED, we note that the hotter companion to this star is an ELM WD of mass $\sim$0.17 M$_{\odot}$, as shown in the upper panel of Figure \ref{Fig.7}. This ELM WD must have formed via case-A/case-B MT, confirming that BMP21 is indeed an FBSS. The cooling age ($\sim$0.03 Gyr) of this WD shown in the lower panel of the same figure suggests that MT has happened very recently. The fact that the star is a rapid rotator also supports a recent MT phenomenon. However, this MT is unlikely to produce the observed chemical enhancements in Sr and Ba. This star is found to be near the separating line between thick disk and halo stars. The large eccentricity of the system may suggest that this is a triple system with an outer sub-luminous companion. This could imply an earlier mass transfer leading to the Sr and Ba enhancement in the FBSS. The large eccentricity can also be due to a possible ejection from a star cluster \citep{khurana2022dynamically}.\\

\textit{BMP29 (BPS CS 22874-0042)}: \cite{preston2000these} classified BMP29 as a RV constant star and a slow rotator (\textit{vsini}= 8 km/s). \cite{hansen2017something} found an $\alpha$-enhancement that is normal for metal-poor halo stars (\big[$\alpha$/Fe\big] = 0.35 $\pm$ 0.09), whereas \cite{preston2000these} found that it was $\alpha$-poor (\big[$\alpha$/Fe\big] $\sim$0.2) and thus a good candidate for an intermediate-age star with extragalactic origin. Slow rotation and RV constancy together could suggest that this may be a face-on system. Regarding the heavy s-process element Ba, \cite{hansen2017something} obtained a good agreement of Ba abundances with solar values. A moderate enhancement for Sr is derived by \cite{hansen2017something}, in contrast to \cite{preston2000these}, who obtained \big[Sr/Fe\big] = --0.4. Its abundance pattern points to a scenario where this star has had time to be enriched by several supernovae at a \big[Fe/H\big] $>$ --2 dex and a mean \big[$\alpha$/Fe\big] of 0.35 dex. This star is more likely enriched by massive rotating stars or normal (massive) supernovae creating the high level of $\alpha$-elements and the low level of s-process elements. \citet{hansen2017something} confirms that this star does not have any companion that could be detected based on both RV and colour. It must have experienced a very different enrichment, possibly from massive stars with inefficient heavy-element production. The parameters obtained from the best-fit SED suggest that the hotter companion is likely to be an LM WD of mass $\sim$0.3 M$_{\odot}$, implying that BMP29 is an FBSS that has formed via an MT. From the WD cooling curves shown in the lower panel of Figure \ref{Fig.7}, it is clear that the age of the WD companion of BMP29 is $\sim$0.15 Gyr, which suggests that MT has happened recently. \\

\textit{ BMP37 (GD 625)}: \cite{preston2000these} classified this star as an SB1 with a period of 84 days and eccentricity 0.07. Based on the enhancement [Ba/Fe] abundance, \cite{sneden2003binary} categorized this star as an FBSS. We note from the best-fit SED parameters that the hotter companion to this star is a normal-mass WD of mass $\sim$0.6 M$_{\odot}$ (refer Figure \ref{Fig.7}). We conclude that BMP37 is an FBSS belonging to the Galactic halo. The WD cooling curve in the same figure suggests that the MT happened $\sim$1.0 Gyr ago, and the star shows moderate rotation. The observed enhancements in \big[Sr/Fe\big] and [Ba/Fe] are in agreement with a normal-mass WD as the hotter companion. GD 625 system is an example of the FBSS with a binary period of 84 days, where a Case-C MT has resulted in the enrichment of Ba.\\

\textit{BMP42 (BPS CS 22166-0041)}: This star is a known SB1 with a period of 486 days and eccentricity of 0.02, a slow rotator (\textit{vsini} = 17 km/s). It is deficient in \big[Sr/Fe\big] as reported by \cite{preston2000these}, although it shows mild enhancement in Ba. From the parameters of the best-fit SED, we discovered a normal-mass WD of mass $\sim$0.6 M$_{\odot}$ associated with this star as the hotter companion, which has formed $\sim$0.1 Gyr ago (Figure~\ref{Fig.7}). Such detection of almost normal-mass WD implies that BMP42 must have formed via Case-C MT, and hence BMP42 is a confirmed FBSS belonging to the halo. This result is in agreement with \cite{preston2000these}, which categorized BMP42 to be an SB1. We note that this system and BMP37 are similar with respect to the Sr and Ba abundances, except that this system has a longer binary period of 486 days.\\

\textit{BMP43 (TYC 8778-1253-1)}: \cite{preston2000these} classified this star as an SB1 with a period of 0.97 days and zero eccentricity suggesting it to be a very short period binary. They also reported it to be a rapid rotator (\textit{vsini}= 60 km/s), deficient in \big[Sr/Fe\big] and \big[Ba/Fe\big]. The best-fitting parameters of the hotter companion suggest that it is likely to be an ELM WD of mass $\sim$0.17 M$_{\odot}$ (upper panel of Fig \ref{Fig.7}). The discovery of ELM WD as a hotter companion supports Case-A/Case-B MT as the possible formation scenario for BMP43 and confirms it to be an FBSS belonging to a thick disk. The Case-A/Case-B MT supports both its SB1 nature as well as the deficiency in \big[Sr/Fe\big] and \big[Ba/Fe\big] known in the literature.\\ 

\textit{BMP46 (BPS CS 22175-0034)}: \cite{preston2000these} classified BMP46 as an RV constant star, a rapid rotator (\textit{vsini} = 60 km/s), and deficient in \big[Sr/Fe\big] and \big[Ba/Fe\big]. The best-fitting SED parameters of the hotter companion suggest that it is an ELM WD of mass $\sim$0.17 M$_{\odot}$) (refer to the upper panel of Figure~\ref{Fig.7}) implying that BMP46 is an FBSS formed via Case-A/Case-B MT and belongs to the thick disk. It is also evident from the lower panel of this figure that the WD must have formed $\sim$1.5 Gyr ago.\\

\textit{BMP49 (BD +24 1676)}: This is a metal-poor (\big[Fe/H\big] = --2.54) star \citep{preston2000these} with a low abundance of [Sr/Fe] and \big[Ba/Fe\big] \citep{zhao2016systematic}. Recently, \cite{smiljanic2021inhomogeneity} studied the kinematics of this BMP star and concluded that it is likely to be a part of the  Gaia Enceladus merger remnant. The parameters from best-fitting SED suggest that the hot companion is a high-mass WD of mass $\sim$0.7 M$_{\odot}$, that has formed $\sim$ 0.31 Gyr ago (Figure~\ref{Fig.7}). We confirm this to be an FBSS and its association with the halo. We also estimated the mass of the WD progenitor to be $\sim$3.0 M$_{\odot}$ using initial $-$ final mass relation of \cite{cummings2018white}. Since the progenitor of the WD is massive, BMP49 is likely to be of intermediate age, though the Gaia Enceladus merger is expected to have happened about 8--11 Gyr ago. This star might have formed from the gas that was part of the parent galaxy. \\

\textit{BMP51 (SDSS J1341+4741)}: BMP51 was identified as an extremely metal-poor (EMP; \big[Fe/H\big] $<$ -- 3.0) star using the high-resolution Hanle Echelle Spectrograph on the 2.0-m Himalayan Chandra Telescope (HCT) \citep{bandyopadhyay2018chemical}. The EMP stars of the Galactic halo are thought to be the direct descendants of the first stars, formed when the universe was only a few hundred million years old, such that their evolution and explosion resulted in the first production of heavy elements \citep{bonifacio2009first}. \cite{bandyopadhyay2018chemical} determined the overall abundance of the $\alpha$-elements to be consistent with the typical halo enhancement of \big[$\alpha$/Fe\big] = 0.4 along with under-abundance of Sr compared to the solar ratio, \big[Sr/Fe\big]= --0.51. Based on the clear under-abundance of the s-process elements, along with its strong carbon over-abundance, they classified BMP51 as CEMP-no star. The parameters from the best-fit SED suggest that the hot companion of BMP51 is likely to be an LM WD of mass $\sim$0.2 M$_{\odot}$ that has formed recently ($\sim$0.31 Gyr ago). The discovery of an LM WD suggests that this star has been formed via Case-A/Case-B MT. We confirm this system to be an FBSS belonging to the galactic halo.

\textit{BMP55 (BPS CS 22966-037)}: This BMP star is an SB1 with a period of 270 days and zero eccentricity, a rapid rotator (\textit{vsini} $\sim$35 km/s), and enhanced in \big[Sr/Fe\big] and \big[Ba/Fe\big] abundances \citep{preston2000these}. The best-fitting parameters of the hot companion suggest it to be an LM WD of mass $\sim$0.2 M$_{\odot}$ (upper panel of Figure~\ref{Fig.7}), which implies that BMP55 is an FBSS formed via Case-A/Case-B MT. It can be noted from the lower panel of Figure~\ref{Fig.6} that the WD companion is $\sim$0.03 Gyr old, suggesting a very recent MT event. This finding is in agreement with the known rapid rotation of the star. However, the enhancements in \big[Sr/Fe\big] and \big[Ba/Fe\big] are possibly either intrinsic to the star or acquired in some other process before MT. This star could belong to either a thick disk or a halo, as the space velocity and the \big[Fe/H\big] values are near the separation of the two populations.\\

\subsection{Final remarks}
To summarize, our study identifies 6 ELM WDs, 3 LM WDs, 2 normal-mass WDs, and 1 high-mass WD as hot companions of 12 BMP stars. The discovery of ELM and LM WDs proves that their corresponding BMP stars were formed through Case-A/Case-B MT. The presence of normal-mass and high-mass WDs as the hot companion, on the other hand, suggests Case-C MT as a possible formation channel of their respective BMP stars. Our findings indicate that MT plays a significant role in the formation of BMP stars, hence confirming that some fraction of BMP stars are indeed FBSS. Nevertheless, we should note that the sample selected for the UVIT observations are either suspected or previously known binaries. Hence, the fraction we confirm as FBSS among the BMP is from this pre-filtered sample. Importantly, the detection and characterisation of the WDs helped to gain insights into their formation mechanisms. 

Out of these 12 BMP stars having WDs as their hot companions, 8 are either known SB1 or binary candidates \citep{preston2000these,hansen2017something,arentsen2019binarity}, 3 are known to be RV constant stars \citep{preston2000these}, and the remaining one have no information on binarity. As we have found WD companions to the 3 RV constant stars, it is likely that the orbit of these binaries may be inclined in a manner so that the RV variations have not been observed and hence classified as RV constant stars. Furthermore, we fitted 10 BMP stars with the single-component SEDs. Seven of these BMP stars are known to be SB1s or binary candidates, two are RV constant stars, and one is an SB2 with an MS star as the companion. For the known SB1s, we suppose that their WD companions may have cooled down and therefore did not show much excess in the UV wavelengths. In the case of BMP2, the binary companion is a known MS star, which is verified in our SED as well. On the other hand, four BMP stars, BMP6, BMP13, BMP14, and BMP48, showed a UV excess but couldn't be fitted with a binary component SED. 

We estimated the masses of all BMP stars that range from 0.96 $-$ 1.42 M$_{\odot}$ by comparing their location with ZAMS displayed in the H-R diagram (Figure~\ref{Fig.7}). Furthermore, the estimated temperatures from SED cover a range from 5250 $-$ 7500 K. We compared the SED-based temperatures of our BMP stars with the BP/RP spectra-based temperatures from Gaia DR3 \citep{babusiaux2022gaia,de2022gaia}. Our SED-estimated temperatures match Gaia DR3 temperatures within 200 K for objects that have single-temperature fits and within 400 K for objects fitted with the binary-component SEDs. We also checked the variability of our BMP stars in Gaia DR3. Four stars, BMP6, BMP37, BMP42, and BMP46 were found to be short time-scale ( $<$ 0.5 $-$ 1 day) MS-type oscillators (Gamma Doradus, Delta Scuti, SX Phoenicis).

\begin{figure}
\includegraphics[width=\columnwidth]{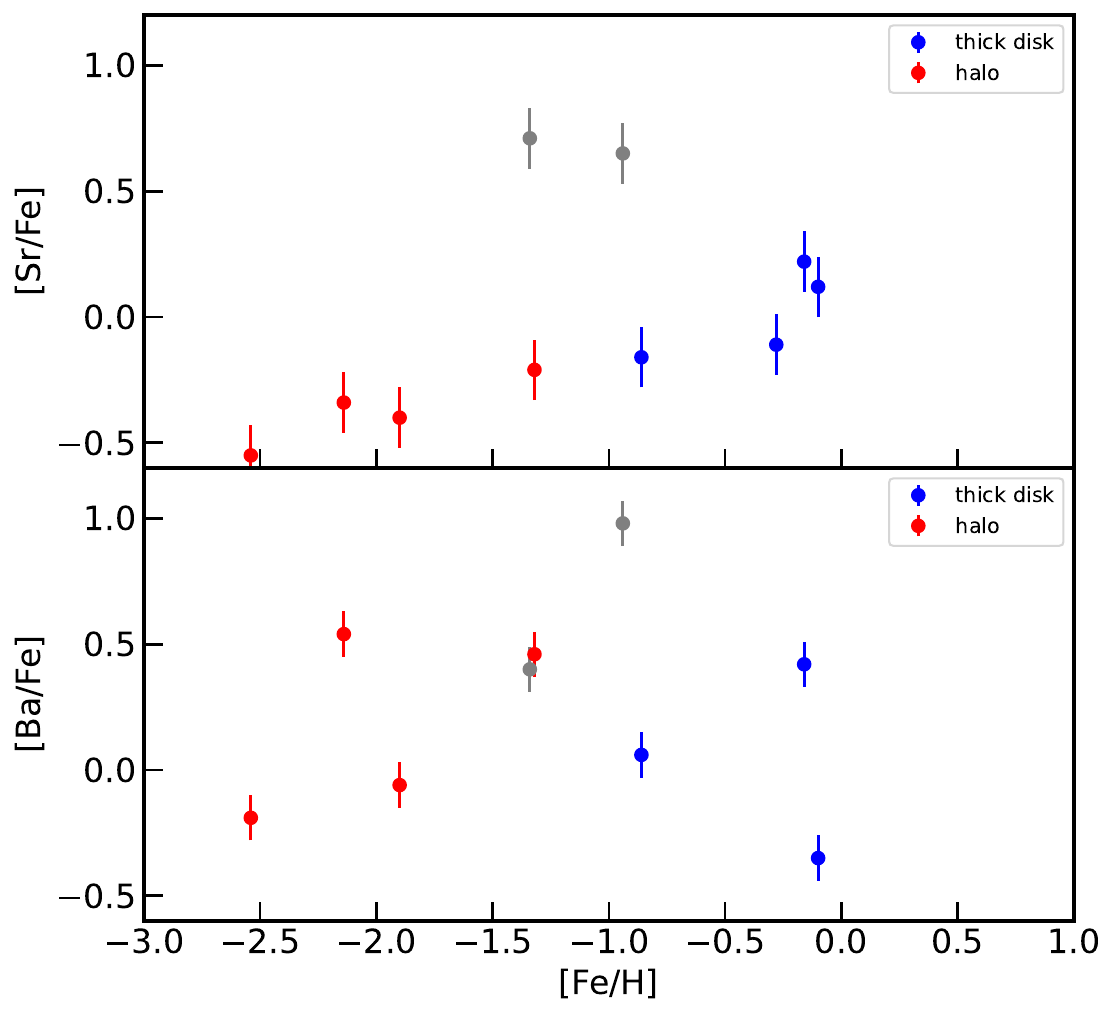}
\caption{The upper panel shows the distribution of \big[Fe/H\big] vs \big[Sr/Fe\big] in all the BMP stars. The lower panel shows the distribution of \big[Fe/H\big] vs \big[Ba/Fe\big] in all the BMP stars. The red dots represent the halo stars, whereas the blue dots represent the thick disk stars. The BMP stars displayed in grey dots do not have any clear distinction. }
\label{Fig.8}
\end{figure}

\cite{smiljanic2007abundance} compared the abundances of normal giants, mild-Ba stars, and Ba stars in order to investigate their origin. In their analysis, they concluded that there is no significant difference in Fe abundances between mild-Ba stars and Ba stars (refer to Figure 12 therein). The two groups share the same \big[Fe/H\big] range as the normal-disk giants under consideration here. Thus, Ba and mild-Ba stars appear to be members of the same stellar population, at least in terms of \big[Fe/H\big] content. We performed a similar analysis on our BMP stars sample and found no correlation between the distribution of \big[Fe/H\big] vs \big[Ba/Fe\big] as shown in Figure~\ref{Fig.8}. Moreover, we have separated our BMP stars sample with blue dots as thick disk stars and red dots as halo stars. BMP55 and BMP21 do not show a clear distinction between these two populations. We have used the abundance values as reported by \cite{preston2000these}, where they considered the mean standard deviation as the errors in the abundances. They estimated error in abundance of Sr to be $\sim$ 0.12, whereas for Ba to be $\sim$0.09.

\cite{jorissen2019barium} studied long-period sample of Ba stars with strong anomalies, hence allowing them to investigate several orbital properties of these post-MT binaries in an unbiased manner. We compared the mass ratio of our Ba enhanced BMP stars, such as BMP3, BMP21, BMP37, and BMP42, with their sample stars of similar Ba content and orbital periods. We found that the mass ratios, and hence the nature of WD companions, are in agreement in both works. However, high-resolution spectroscopic studies of the above mentioned BMP stars are important to get the abundances of other elements as well.

It is believed that the Gaia-Enceladus-Sausage was the last major merger of our Galaxy and happened about 8$-$11 Gyr ago \citep{vincenzo2019fall, grand2020dual}. This gas-rich merger is expected to have triggered an intense star formation, during which most canonical thick disk stars formed. BMP49 (BD +24 1676), is a likely member of the Gaia-Enceladus-Sausage as found by \cite{smiljanic2021inhomogeneity}. We detect a massive WD as the hot companion to this star, suggesting an intermediate age. This raises a possibility that this star might be formed from the gas of the parent galaxy, after merging with the Milky Way. 

We also have examined if there is any correlation between abundances of Li, s-process elements, and $\alpha$ elements of BMP stars with the nature of their corresponding WD companions; however, no clear correlation was observed. Furthermore, not all BMP stars that we suggest as FBSS have depletion in Li abundance. Similarly, no depletion of Li has been seen in post-MT systems in M67 \citep{hobbs1991observed,lombardi2002stellar}. We have listed the values of element abundances from literature in Table~\ref{Table1} and Table~\ref{Table2}. It is noteworthy that accurate determination of abundances of elements such as Li, Mg, Ca, C, O, Sr, and Ba is very useful to establish the connection between the chemical peculiarities of BMP stars and the nature of their WD companions. The element abundances present in the literature are either incomplete or determined long ago using low-resolution spectra. Therefore, it is crucial to determine the element abundances of BMP stars with more accuracy.

\section{Summary and conclusions} \label{Section 5} 

We summarize the main findings from this paper in this paper as follows:

\begin{enumerate}

\item With the aim of detecting and characterising FBSS, we studied 27 BMP stars (that are potential binaries) using \textit{Astrosat}/UVIT images in two FUV filters, F148W and F169M. Based on the multi-wavelength SEDs fitted to all BMP stars, 17 showed the excess in UV flux $>$ 50$\%$, whereas 10 of them showed an excess $<$ 50$\%$. Out of 17 BMP stars showing excess in UV fluxes, we could successfully fit the binary component SEDs to 13 of them. In the case of the remaining 4 BMP stars, we noted that either no models were fitting with the observed fluxes or the best-fitting SEDs were giving unreliable temperature values of the hot companions. \\

\item We report the first-ever discovery of hot companions of 12 BMP stars (6 ELM WDs, 3 LM WDs, 2 normal-mass WDs, and 1 high-mass WD). The temperatures of these WDs range from 10,500 $-$ 40,000 K, luminosities vary from 0.01 $-$ 0.67 L$_{\odot}$, and radii vary from 0.01 $-$ 0.22 R$_{\odot}$. The estimated temperatures of BMP stars vary from 6500 $-$ 7750K, and their masses have a range of 0.96 $-$ 1.42 M$_{\odot}$. In the case of the remaining BMP star fitted with the double component SED, the binary companion is found to be an MS star.  \\

\item We note that out of 12 BMP stars with WDs as hot companions, 8 are either known to be SB1 or binary candidates. We confirmed their binarity by detecting hot companions to them. The remaining 3 BMP stars are known to be RV constant stars. We suggest that the orbits of these binary systems might be of low inclination.\\

\item The discovery of ELM and LM WDs as the hot companions of BMP stars validates the Case-A/Case-B MT as their formation channel, whereas the discovery of normal-mass and high-mass WDs signifies Case-C MT as the formation mechanism of the associated BMP stars. We conclude that at least 12 out of 27 (44 $\%$) BMP stars are FBSS formed via the MT formation channel. \\

\item We fitted single-component SEDs to 10 BMP stars with $<$ 50$\%$ excess in UV flux. Out of these 10 BMP stars, 7 are known to be SB1s, 2 are RV constant stars, and 1 is a SB2. In the case  of known SB1s, the WD companions may have cooled down beyond the detection limit of UVIT.\\

\item We also calculated the space velocities of all the BMP stars and grouped them into a thick disk and halo stars. In the studied sample, all 5 BMP stars in the thick disk have ELM WDs as their hot companions. Two BMP stars are intermediate between the halo and thick disk, and these have ELM/LM WDs. In the case of 5 halo BMP stars, we detect LM, normal-mass, and high-mass WDs as their hot companions. \\

\item This study confirms that GD 625 (BMP37) and BPS CS 22166-0041 (BMP42) are FBSS with a WD companion belonging to the Galactic halo and showing enhancement in [Ba/Fe]. On the other hand, we confirm BPS CS 22874-0042 (BMP29) to be an FBSS, with an LM WD companion located in the Galactic halo. This study also finds that BD +24 1676 (BMP49), belonging to the Gaia Enceladus merger remnant, is an FBSS. \\

\item We also checked if there is any correlation between the chemical peculiarities of FBSS, their hot companions and the orbital properties. The periods are $<$ 1000 days, with two having large eccentricity. Except one, the FBSS of thick disk do not show significant enhancement in \big[Sr/Fe\big] and \big[Ba/Fe\big]. This is in agreement with Case-A/Case-B MT. However, no clear trend was observed among the halo FBSS. As the halo stars may have different origins, it is important to disentangle the primordial abundance from that acquired through accretion. Detailed abundance analysis of these FBSS using high-resolution spectroscopy is required to estimate their chemical properties. 
\end{enumerate}

\section*{Acknowledgements}

We thank the anonymous referee for the valueable suggestion/corrections, which has helped in improving the quality of the paper.
This work uses the data from UVIT onboard the \textit{AstroSat} mission of the Indian Space Research Organisation (ISRO). UVIT is a collaborative project between various institutes, including the Indian Institute of Astrophysics (IIA), Bengaluru, The Indian-University Centre for Astronomy and Astrophysics (IUCAA), Pune, Tata Institute of Fundamental Research (TIFR), Mumbai, several centres of Indian Space Research Organisation (ISRO), and Canadian Space Agency (CSA). AS thanks for the support of the SERB power fellowship. VJ thanks the Alexander von Humboldt Foundation for their support. This work has made use of TOPCAT \citep{taylor2011topcat}, Matplotlib \citep{hunter2007matplotlib}, IPython \citep{perez2007ipython}, Scipy \citep{oliphant2007scipy,millman2011python} and Astropy, a community-developed core Python package for Astronomy \citep{price2018astropy}. This publication also makes use of VOSA, developed under the Spanish Virtual Observatory project supported by the Spanish MINECO through grant AyA2017-84089. 

\section*{Data availability}

The data underlying this article are publicly available at \url{https://astrobrowse.issdc.gov.in/astro_archive/archive/Home.jsp} The derived data generated in this research will be shared on reasonable request to the corresponding author.

\bibliographystyle{mnras}
\bibliography{references}

\begin{table*}
\caption{Parameters from the literature of BMP stars fitted with the single-component SEDs. For each of them, coordinates in Columns 2 -- 3, distance in Column 4, extinction in Column 5, metallicity in Column 6, a period in Column 7, eccentricity in Column 8, whether radial velocity constant or binary star in Column 9, radial velocity in Column 10, rotational velocity in Column 11, Sr and Ba abundances in Columns 12 --13.}
\addtolength{\tabcolsep}{-3pt}
\small
\begin{tabular}{cccccccccccccc}
\\
\hline
Name&RA&DEC&D&A$_{v}$&[Fe/H]&P&e&Binarity&RV&vsini&[Sr/Fe]&[Ba/Fe]\\
\\
\hline
BMP6&325.676875&$-$54.31194&6826.95$\pm$960.33&0.051$\pm$0.001&$-$4.02&300&-&Binary&102.18$\pm$10.07&-&$-$0.55&$-$0.33\\
(HE 2139-5432)&&&&&&&\\
\hline
BMP10&326.87183&$-$39.42213&1251.39$\pm$30.58&0.074$\pm$0.002&$-$1.63&-&-&RVC&$-$49.80$\pm$0.86&12&$-$0.04&0.43&\\
(BPS CS 22948-0079)&&&&&&&&&&\\
\hline
BMP11&304.97400&$-$38.94338&2226.29$\pm$128.13&0.155$\pm$0.003&$-$1.36&174&0.50&SB1&$-$47.80$\pm$4.06&45&0.66&0.15&\\
(BPS CS 22885-0048)&&&&&&&&&\\
\hline
BMP13&354.53260&$-$35.88117&1346.87$\pm$45.02&0.039$\pm$0.001&$-$0.14&710&0&SB1&3.60$\pm$4.37&55&$-$0.41&$-$0.55&\\
(BPS CS 22941-0035)&&&&&&&&\\
\hline
BMP14&352.47016&$-$35.21788&1300.69$\pm$59.11&0.049$\pm$0.003&$-$2.43&324&0.2&SB1&$-$49.00$\pm$4.34&25&$-$0.43&0.34\\
(BPS CS 22941-0005)&&&&&&&\\
\hline
BMP15&358.83895&$-$34.80058&894.70$\pm$19.89&0.037$\pm$0.001&$-$1.88&302.5&0.12&SB1&67.20$\pm$6.42&15&0.27&$-$0.05&\\
(BPS CS 22876-0008)&&&&&&&&&&\\
\hline
BMP17&0.26576&$-$33.80442&1866.25$\pm$82.44&0.034$\pm$0.001&$-$1.1&176.9&0.1&SB1&$-$73.00$\pm$5.42&12&0.73&0.5&\\
(BPS CS 22876-0021)&&&&&&&&&&&\\
\hline
BMP20&352.12387&$-$32.81308&649.37$\pm$6.47&0.040$\pm$0.002&$-$2.03&-&-&RVC&$-$122.40$\pm$0.66&10&0.19&$-$0.21&\\ 
(BPS CS 22941-0012)&&&&&&&&&&&&\\
\hline
BMP23&12.56616&$-$30.99916&947.85$\pm$62.55&0.059$\pm$0.002&$-$2.02&194&0.35&SB1&74.20$\pm$2.87&10&0.82&1.33&\\
(CD 31-306)&&&&&&&&&&&\\
\hline
BMP30&10.19987&$-$24.12666&518.65$\pm$7.09&0.044$\pm$0.001&$-$3&346&0.3&SB1&-&-&$-$1.2&$-$0.67\\
(CD 24-266)&&&&&&&&&&\\
\hline
BMP36&7.29483&$-$19.16883&1108.42$\pm$33.19&0.043$\pm$0.001&$-$3.32&-&-&SB2&47.13$\pm$0.42&-&0.34&-&\\
(BPS CS 29527-0015)&&&&&&&&&&&\\
\hline
BMP44&13.04783&$-$11.07783&1408.409$\pm$41.35&0.08$\pm$0.010&$-$1.3&29.61&0.35&SB1&$-$95.02$\pm$10.32&120&-&-&\\
(BPS CS 22166-0004)&&&&&&&&&&\\
\hline
BMP48&5.15104&23.79361&1049.07$\pm$64.34&0.080$\pm$0.003&$-$2.95&-&-&?&184.41$\pm$0.30&-&$-$0.02&$-$0.31&\\
(BPS BS 17570-0063)&&&&&&&\\
\hline
BMP50&8.18087&24.22233&323.522$\pm$8.22&0.06$\pm$0.002&$-$0.91&840.4&0.168&SB1&36.35$\pm$0.94&30&-&-&\\
(BD +23 74)&&&&&&&&\\
\hline
\label{Table1}
\end{tabular}
\end{table*}

\begin{table*}
\caption{Parameters from literature of BMP stars fitted with the double-component SEDs. All the Columns are similar as in Table \ref{Table1}.}
\addtolength{\tabcolsep}{-3pt}
\small
\begin{tabular}{cccccccccccccc}
\hline
Name&RA&DEC&D&A$_{v}$&[Fe/H]&P&e&Binarity&RV&vsini&[Sr/Fe]&[Ba/Fe]\\
\\
\hline
BMP2 &301.48008&$-$59.28663&1156.44$\pm$22.81&0.093$\pm$0.001&$-$3.4&19.16&0.26&SB2&243.00$\pm$2.00&10&$-$1.1&-&\\
(BPS CS 22873-0139)&&&&&&&&&&&&\\
\hline 
BMP3&295.93700&$-$56.16911&972.28$\pm$32.09&0.124$\pm$0.003&$-$0.16&1.23&0&SB1&32.80$\pm$	15.81&60&0.22&0.42&\\
(TYC 8778-1253-1)&&&&&&&&&&&&\\
\hline
BMP4&298.26675&--55.73283&1336.59$\pm$37.52&0.142$\pm$0.005&--0.86&-&-&RVC&74.60$\pm$0.82&12&$-$0.16&0.06&\\
(BPS CS 22896-0173)&&&&&&&&&&&&\\
\hline
BMP5&293.31445&$-$54.98111&2008.55$\pm$93.03&0.142$\pm$0.002&$-$0.1&31.66&0.45&Binary &85.00$\pm$4.37&45&0.12&$-$0.35&\\
(BPS CS 22896-0103)&&&&&&&&&&&&\\
\hline
BMP21&19.30054&$-$32.44944&1541.89$\pm$40.92&0.062$\pm$0.010&$-$0.94&64.7&0.44&SB1&$-$69.21$\pm$3.96&40&0.65&0.98&\\
(BPS CS 29518-0024)&&&&&&&&&&&&\\
\hline
BMP29&219.50775&$-$24.97908&766.02$\pm$9.93&0.235$\pm$0.005&$-$1.9&-&-&RVC&176.20$\pm$0.72&8&$-$0.4&$-$0.06\\
(BPS CS 22874-0042)&&&&&&&&&&&&\\
\hline
BMP37&8.96233&$-$17.95005&1140.59$\pm$34.48&0.05$\pm$0.001&$-$2.14&84&0.07&SB1&$-$185.80$\pm$9.97&35&$-$0.34&0.54\\
(GD 625)&&&&&&&&&&&&\\
\hline
BMP42&17.73504&$-$13.71205&1092.57$\pm$41.65&0.062$\pm$0.003&$-$1.32&486&0.024&SB1&36.90$\pm$8.17&-&$-$0.21&0.46\\
(BPS CS 22166-0041)&&&&&&&&&&&&\\
\hline
BMP43&12.78075&$-$11.14219&788.83$\pm$17.03&0.08$\pm$0.001&$-$0.68&0.97&0&SB1&95.02$\pm$10.32&60&$-$0.74&-&\\
(TYC 8778-1253-1)&&&&&&&&&&&&\\
\hline
BMP46&35.08962&$-$10.63588&716.38$\pm$7.55&0.07$\pm$0.001&$-$0.28&-&-&RVC&27.21$\pm$0.73&-&$-$0.11&-&\\
(BPS CS 22175-0034)&&&&&&&&&&&&\\
\hline
BMP49&112.67270&24.08516&250.08$\pm$0.93&0.040$\pm$0.002&$-$2.54&-&-&Binary&$-$237.09$\pm$0.48&-&$-$0.55&$-$0.19\\
(BD +24 1676)&&&&&&&&&&&&\\
\hline
BMP51&205.43574&47.69068&431.59$\pm$3.46&0.046$\pm$0.002&$-$3.2&-&-&-&-209.83$\pm$0.71&-&$-$0.51&$-$0.73\\
(SDSS J1341+4741)&&&&&&&&&&&&\\
\hline
BMP55&355.91075&$-$31.80510&1520.21$\pm$45.46&0.040$\pm$0.001&$-$1.34&270&0&SB1&59.01$\pm$5.09&35&0.71&0.4\\
(BPS CS 22966-037)&&&&&&&&&&&&\\
\hline
\label{Table2}
\end{tabular}
\begin{tablenotes}
\item {Distance (D) : \cite{bailer2018estimating}}
\item {Extinction (A$_{v}$) : \cite{preston2000these} and Galactic dust reddening and extinction maps.}
\item {Metallicity([Fe/H]) : \cite{wilhelm1999spectroscopy,preston2000these,hansen2017something,arentsen2019binarity}} 
\item {Period (P) : \cite{preston2000these,sneden2003binary,hansen2017something,arentsen2019binarity} } 
\item {Eccentricity (e) : \cite{preston2000these,sneden2003binary,hansen2017something,arentsen2019binarity}} 
\item {Binarity : \cite{preston2000these,yong2012most,hansen2017something,arentsen2019binarity} } 
\item {Radial velocity (RV): \cite{preston2000these,yong2012most,hansen2017something,arentsen2019binarity}} 
\item {Rotational velocity (vsini) : \cite{preston2000these,sneden2003binary,hansen2017something,arentsen2019binarity}}
\item{[Sr/Fe] and [Ba/Fe]: \cite{preston2000these}, \cite{bonifacio2009first}, \cite{yong2012most}}
 
\end{tablenotes}
\end{table*}

\begin{table*}
\centering
\caption{\label{Table3} For each BMP star, UVIT F148W flux in Column 2, GALEX FUV flux in Column 7, UVIT F169M flux in Column 4, GALEX NUV flux in Column 5, PAN-STARRS and \textit{Gaia} DR3 fluxes in Columns 6 --13, 2MASS J, H, and Ks fluxes in Columns 14--16, and WISE W1, W2, W3, and W4 fluxes in Columns 17--20. All flux values are listed in the unit of erg s$^{-1}$ cm$^{-2}$ \AA$^{-1}$. The full table is available in online version.}

\begin{tabular}{ccccccccc}
\hline
\\
Name&UVIT.F148W$\pm$err&GALEX.FUV$\pm$err&UVIT.F169M$\pm$err&GALEX.NUV$\pm$err&\\
PS1.g$\pm$err&GAIA3.Gbp$\pm$err&GAIA3.G$\pm$err&PS1.r$\pm$err&PS1.i$\pm$err&\\
GAIA3.Grp$\pm$err&PS1.z$\pm$err&PS1.y$\pm$err&2MASS.J$\pm$err&2MASS.H$\pm$err&\\
2MASS.Ks$\pm$err&WISE.W1$\pm$err&WISE.W2$\pm$err&WISE.W3$\pm$err&WISE.W4$\pm$err\\

\hline
\\
BMP2&4.389e-16 $\pm$ 7.168e-18& 1.562e-16 $\pm$ 4.304e-17&5.535e-16 $\pm$ 5.995e-18&6.898e-15 $\pm$	1.028e-16&\\
&1.182e-14 $\pm$ 3.265e-17&8.822e-15 $\pm$ 2.250e-17&-&-&\\
6.291e-15 $\pm$ 2.232e-17&-&-&2.301e-15 $\pm$ 5.723e-17&1.070e-15 $\pm$ 2.759e-17&\\
3.851e-16 $\pm$ 1.029e-17&8.153e-17 $\pm$ 1.727e-18&
2.357e-17 $\pm$ 5.210e-19&9.030e-19 $\pm$ 2.653e-19&1.632e-18 $\pm$ 0.000e+00\\
\\
\hline
\\
BMP3&3.062e-14 $\pm$ 9.046e-15&2.466e-14 $\pm$ 5.273e-16&2.721e-14 $\pm$ 2.399e-15&4.050e-14 $\pm$	2.457e-16&\\
&-&-&-&-&\\
-&-7.758e-15 $\pm$	1.643e-16&-&-&3.016e-15 $\pm$ 6.666e-17&\\1.124e-15 $\pm$	2.381e-17& 
2.230e-16 $\pm$ 4.518e-18&6.460e-17 $\pm$ 1.249e-18&1.442e-18 $\pm$ 2.815e-19&1.753e-18 $\pm$ 0.000e+00\\
\\
\hline 
\end{tabular}
\end{table*}

\begin{table*}
\caption{The best-fit parameters of BMP stars fitted with the single-component SEDs. For each of them, we have listed  luminosity, temperature, and radius in Columns 2 $-$ 4, the reduced $\chi^{2}_{r}$ values in Column 5, the scaling factor in Column 6, the number of data points used to fit the SED is given in Column 7, and the values of vgf$_{b}$ parameter in Column 8.} 
	\adjustbox{max width=\textwidth}{
	\begin{tabular}{ccccccccccc}
		\hline
		\\
		Name&Luminosity&T$\mathrm{_{eff}}$&Radius&$\chi^{2}_{r}$&Scaling factor&N$_{fit}$&$vgf_{b}$\\
	    ~&~&[L$_\odot$]&[K]&[R$_\odot$]&&~&~&
		 \\
		\hline 
		\\

BMP6&40.01$\pm$11.13&5500$\pm$125&7.66$\pm$1.07&40.65&6.40E-22&14&0.41\\
BMP10&4.65$\pm$0.23&6750$\pm$125&1.57$\pm$0.03&2.85&8.06E-22&15&0.21\\
BMP11&9.25$\pm$1.08&6750$\pm$125&2.22$\pm$0.12&3.01&5.07E-22&15&0.86\\
BMP13&3.48$\pm$0.48&7000$\pm$125&1.26$\pm$0.04&3.07&4.51E-22&14&0.18\\
BMP14&2.22$\pm$0.20&7000$\pm$125&0.94$\pm$0.04&6.80&2.69E-22&14&0.31\\
BMP15&1.82$\pm$0.08&7000$\pm$125&0.92$\pm$0.02&3.07&5.40E-22&15&0.13 \\
BMP17&4.74$\pm$0.43&7000$\pm$125&1.48$\pm$0.06&2.88&3.23E-22&15&0.38 \\
BMP20&3.93$\pm$0.00&7000$\pm$125&1.34$\pm$0.00&1.90&2.18E-22&15&0.09 \\
BMP23&9.14$\pm$1.22&7000$\pm$125&1.92$\pm$0.12&36.62&2.08E-22&18&0.63\\
BMP30&2.06$\pm$0.05&7000$\pm$125&0.97$\pm$0.01&1.76&1.81E-22&14&0.38 \\
BMP36&2.43$\pm$0.14&6500$\pm$125&1.24$\pm$0.03&140.8&6.41E-22&16&0.37\\
BMP44&9.81$\pm$0.59&8000$\pm$125&1.64$\pm$0.04&10.60&6.97E-22&19&0.78 \\
BMP48&1.74$\pm$0.21&6250$\pm$125&1.12$\pm$0.06&10.43&5.82E-22&18&1.12\\
BMP50&1.04$\pm$0.54&7500$\pm$125&10.40$\pm$0.54&6.94&1.77E-22&14&0.12\\

\hline
		\label{Table4}
\end{tabular}
}

\end{table*}

\begin{table*}
\caption{The best-fit parameters of BMP stars fitted with the double-component SEDs. For each of them, whether cooler (A) or hotter (B) companion in Column2, luminosity, temperature, and radius in Columns 3 $-$ 5, the reduced $\chi^{2}_{r}$ values in Column 6 (the $\chi^{2}_{r}$ values of the single fits are given in the brackets), scaling factor in Column 7, number of data points used to fit the SED is given in Column 8, and the values of $vgf_{b}$ parameter in Column 9 (the $vgf_{b}$ values of the single fits are given in the brackets).}

	\adjustbox{max width=\textwidth}{
	\begin{tabular}{cccccccccccc}
		\hline
		\\
Name&Component&Luminosity&T$\mathrm{_{eff}}$&Radius&$\chi^{2}_{r}$&Scaling factor&N$_{fit}$&$vgf_{b}$\\
	    ~&~&~&[L$_\odot$]&[K]&[R$_\odot$]&&~&~&
        \\
		\hline 
		\\
BMP2&A&1.29$\pm$0.05&5750$\pm$125&1.14$\pm$0.02&14.58&4.971E-22&12&0.51\\
&B&2.60$\pm$0.11&7000$\pm$125&1.09$\pm$0.02&-&4.574E-22&-&-\\		
		
BMP3&A&13.88$\pm$0.96&7750$\pm$125&1.69$\pm$0.05&79.12 (199.7)&1.54E-21&13&0.79 (13.4)\\
&B&1.29$^{+0.21}_{-0.19}$&16750$\pm$250&0.14$\pm$0.01&-&8.87E-24&-&\\

BMP4&A&3.83$\pm$0.22&7000$\pm$125&1.33$\pm$0.03&54.54 (92.98)&5.05E-22&13&2.97 (3.13)\\
&B&0.04$^{+0.01}_{-0.01}$&13000$\pm$250&0.04$\pm$0.00&-&4.76E-25&-&\\

BMP5&A&5.75$\pm$0.54&8000$\pm$125&1.24$\pm$0.07&157.93 (168.5)&1.94E-22&12&3.42 (4.65)\\
&B&0.14$^{+0.03}_{-0.03}$&12250$\pm$250&0.08$\pm$0.00&-&5.75E-24&-&\\

BMP21&A&3.04$\pm$0.16&7500$\pm$125&1.03$\pm$0.02&13.01 (44.77)&1.77E-22&13&1.57 (2.12)\\
&B&0.13$^{+0.02}_{-0.02}$&17250$\pm$250&0.04$\pm$0.00&-&1.10E-22&-&-\\

BMP29&A&1.73$\pm$0.04&6250$\pm$125&1.12$\pm$0.01&5.33 (41.05)&1.09E-21&15&0.22 (0.31)\\
&B&0.01$^{+0.01}_{-0.00}$&12500$\pm$250&0.02$\pm$0.00&-&8.11E-21&-&- \\

BMP37&A&2.78$\pm$0.03&7250$\pm$125&1.05$\pm$0.00&40.71 (17.42)&4.34E-22&16&0.19 (3.08)\\
&B&0.03$^{+0.02}_{-0.01}$&19750$^{+1250}_{-1000}$&0.01$\pm$0.00&-&8.13E-26&-&\\

BMP42&A&2.02$\pm$0.15&6750$\pm$125&0.91$\pm$0.03& 6.18 (91.72)&3.52E-22&20&0.15 (2.43) \\
&B&0.02$^{+0.01}_{-0.01}$&17500$\pm$250&0.01$\pm$0.00&-&6.88E-26&-&-\\

BMP43&A&11.01$\pm$0.45&7750$\pm$125&1.76$\pm$0.03&58.39 (133.72)&2.54E-21&14&3.67 (7.12)\\
&B&0.57$^{+0.12}_{-0.10}$&13500$\pm$250&0.14$\pm$0.01&-&1.73E-23&-&-\\

BMP46&A&3.88$\pm$0.08&6750$\pm$125&1.44$\pm$0.01&43.46 (108.51)&2.06E-21&15&0.4 (5.54)\\
&B&0.08$^{+0.01}_{-0.01}$&11000$\pm$250&0.08$\pm$0.00&-&5.07E-24&-&-\\

BMP49&A&2.94$\pm$0.04&6750$\pm$125&1.25$\pm$0.00&2.31 (47.37)&1.28E-20&10&0.12 (3.87) \\
&B&0.01$^{0+.00}_{-0.00}$&18500$^{+250}_{-500}$&0.01$\pm$0.00&-&2.96E-24&-&-\\

BMP51&A&4.94$\pm$0.09&6250$\pm$125&1.92$\pm$0.01& 150.3 (430.8) &1.00E-20&17&0.55 (2.90)\\
&B&0.02$^{+0.00}_{-0.00}$&13000$^{+250}_{-250}$&0.03$\pm$0.00&-&1.27E-20&-&-\\

BMP55&A&4.75$\pm$0.29&7250$\pm$125&1.38$\pm$0.04& 4.46 (513.43) &4.20E-22&14&0.27 (1.87)\\
&B&0.06$^{+0.02}_{-0.01}$&17250$^{+250}_{-500}$&0.03$\pm$0.00&-&1.90E-25&-&-\\

\hline
		\label{Table5}
\end{tabular}
}
\end{table*}

\begin{table*}
\caption{Parameters of thick disk stars fitted with the double-component SED. For each of them, we have listed metallicity in Column 2, period in Column 3, eccentricity in Column 4, Sr and Ba content in Columns 5-6, rotational velocity in Column 7, space velocity in Column 8, mass of BMP star in Column 9, mass of WD in Column 10, age of WD in Column 11, and the nature of WD in Column 12. }
\begin{tabular}{ccccccccccccc}
\hline
\hline
Name&[Fe/H]&P&e&[Sr/Fe]&[Ba/Fe]&vsini&Space velocity&Mass (star)&Mass (WD)& Age (WD)&Nature of WD\\
\hline  
&(dex)&(days)&&&&(km/s)&(km/s)&(M$_{\odot}$)&(M$_{\odot}$)&(Gyr)&\\
\hline
\\

BMP3&-0.16&1.23&0&0.22&0.42&60&68.5&1.42&0.16&-&ELM\\
BMP4&-0.86&-&-&-0.16&0.06&12&79.3&1.12&0.18&0.31&ELM\\
BMP5&-0.1&31.66&0.45&0.12&-0.35&45&88.1&1.20&0.16&1.0&ELM\\
BMP43&-0.68&0.97&0&-0.74&-&60&52.5&1.36&0.17&-&ELM\\
BMP46&-0.28&-&-&-&0.11&-&45.9&1.15&0.17&1.5&ELM\\

\hline
\label{Table6}
\end{tabular}
\end{table*}

\begin{table*}
\caption{Parameters of halo stars fitted with the double-component SEDs. The Columns are similar as mentioned in Table \ref{Table6}.}
\begin{tabular}{cccccccccccccc}
\hline
\hline
Name&[Fe/H]&P&e&[Sr/Fe]&[Ba/Fe]&vsini&Space velocity&Mass (star)&Mass (WD)&Age (WD) &Nature of WD\\
\hline  
&(dex)&(days)&&&&(km/s)&(km/s)&(M$_{\odot}$)&(M$_{\odot}$)&(Gyr)\\
\hline
\\
BMP29&-1.9&-&-&-0.4&-0.06&8&282.7&0.97&0.3&0.15&LM\\
BMP37&-2.14&84&0.07&-0.34&0.54&35&382.7&1.07&0.6&0.1&Normal-mass\\
BMP42&-1.32&486&0.024&-0.21&0.46&-&245.7&0.96&0.6&0.1&Normal-mass\\
BMP49&-2.54&-&-&-0.55&-0.19&-&336.1&1.07&0.7&0.31&High-mass\\
BMP51&-3.2&-&-&-0.51&-0.73&-&378.1&-&-&0.31&LM\\
\hline
\label{Table7}
\end{tabular}
\end{table*}

\appendix

\section{Data reduction using CCDLAB}
\label{Appendix}

The steps taken to perform the reduction of L1 data to obtain science-ready images are as follows: \\

1) Extraction of L1 data: Generally, the targets are observed over multiple orbits. Therefore, the L1 data products consist of a compressed archive file containing multiple sets of data organized by individual orbits in the form of FITS binary tables. To access this archive file, we first extract the data. Then the digestion of the data and drift correction is performed in an automated mode. This is achieved by clicking the \textit{Extract L1 gz or zip archives} task.\\
2) Registration of images: The drift-corrected images still suffers from orbit-wise translational and rotational shift. In order to eliminate this effect, we perform the registration of all the images  by selecting two bright sources in the image using the \textit{General registration} task. \\
3) Merging: After the alignment of data via registration, we merge the orbit-wise data into the master file using the \textit{Merge centroid list} task. \\
4) Optimization of point spread function (PSF): The merged images exhibit minor flaws caused by the disparity in data sampling rates between the VIS data, sampled at 1 Hz, and the UV data, sampled at 29 Hz. Additionally, a slight variation in pointing between the VIS and FUV telescopes is observed due to thermal stick-slip \citep{postma2021uvit}. To address these issues, we perform the optimization of the PSF using \textit{optimizing the PSF} task. This task involved stacking images of bright sources with exposures of $\sim$20 s, resulting in a minor correction for any drift and optimizing the overall PSF. \\
5) Applying World Coordinate Solution (WCS): We utilise the coordinate matching algorithm using the \textit{Auto WCS} task within CCDLAB. It first extracts the bright sources within UV images and takes bright \textit{Gaia} G$_{BP}$ sources for reference. Then it compares the two catalogues by applying least square solution. Finally, the task applies the WCS solution to all the sources which is also added in the header files of the images.\\
6) Finalizing the science products: The final stage involves preparing the science image files for distribution to end-users and performing a cleanup process to remove all intermediate processing folders and files from the computer system. We utilise \textit{Finalize science products} task to achieve this.

\section{Photometry using curve-of-growth technique in CCDLAB}
\label{Appendix2}

This method is best suited for clean isolated sources such as the field stars under study. We use the \textit{COG} task under\textit{PSE} (point source extractor) function in CCDLAB. It creates a circle around the source of interest with the user defined radius in pixels. The option \textit{view COG} gives a number of pixels vs number of counts plot in which we provide the number of points to be fitted to this curve. The slope of the line that fits this curve gives the background counts per pixel, whereas the intercept gives the source counts per pixel. In order to get the number of counts per pixel per second (CPS), we divide the counts per pixel values with the exposure time of the image. Then we convert this CPS into magnitudes, which in turn are converted to the fluxes, using the zero points and unit conversions of UVIT filters as given in \cite{tandon2020additional}. Similarly, we use the errors in CPS to obtain the errors in FUV fluxes. This method of performing photometry using CCDLAB has been used in previous  studies such as \cite{leahy2022astrosat}.

\section{Additional Figures}

\begin{figure*} 
\includegraphics[scale=0.25]{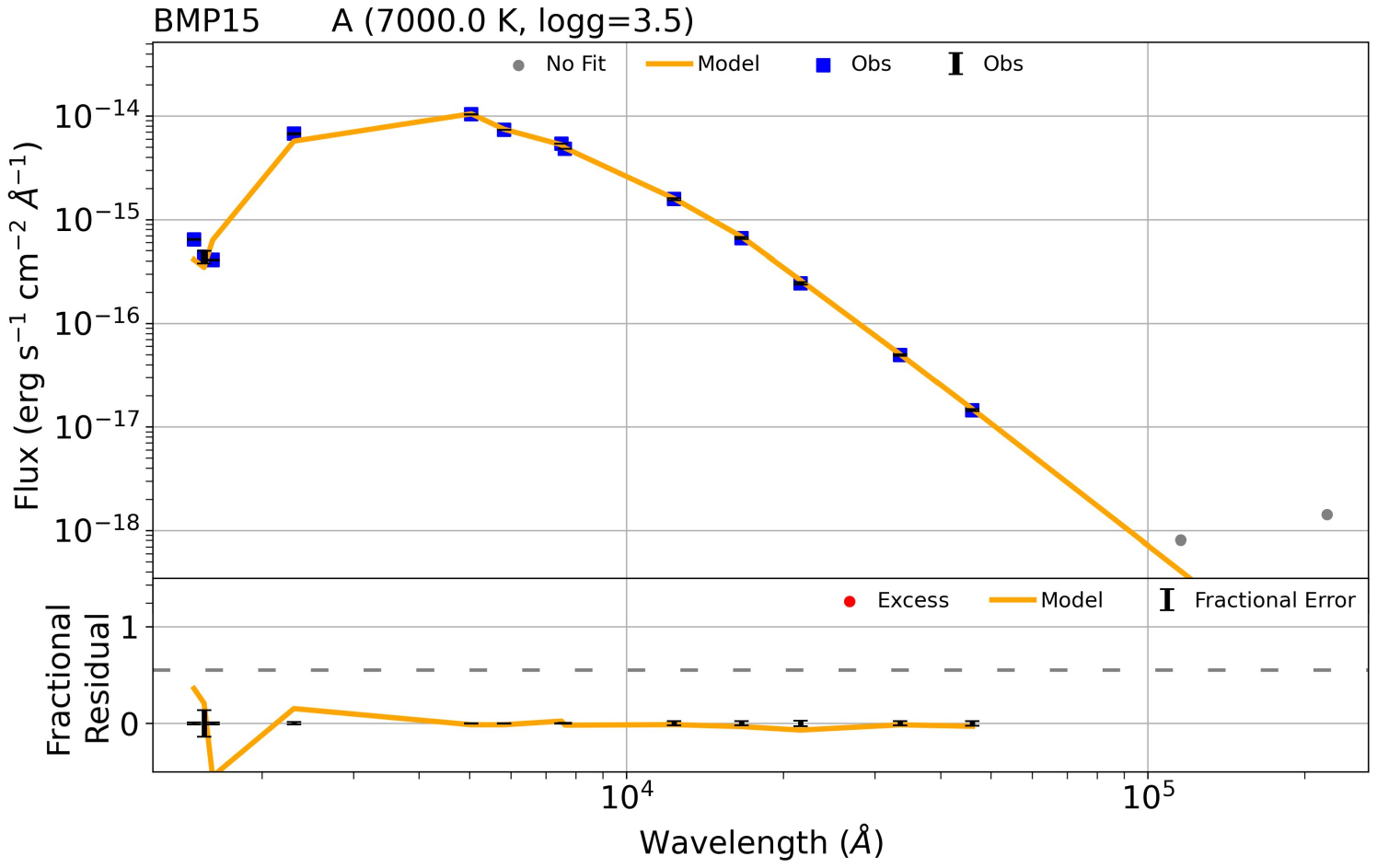} 
\includegraphics[scale=0.25]{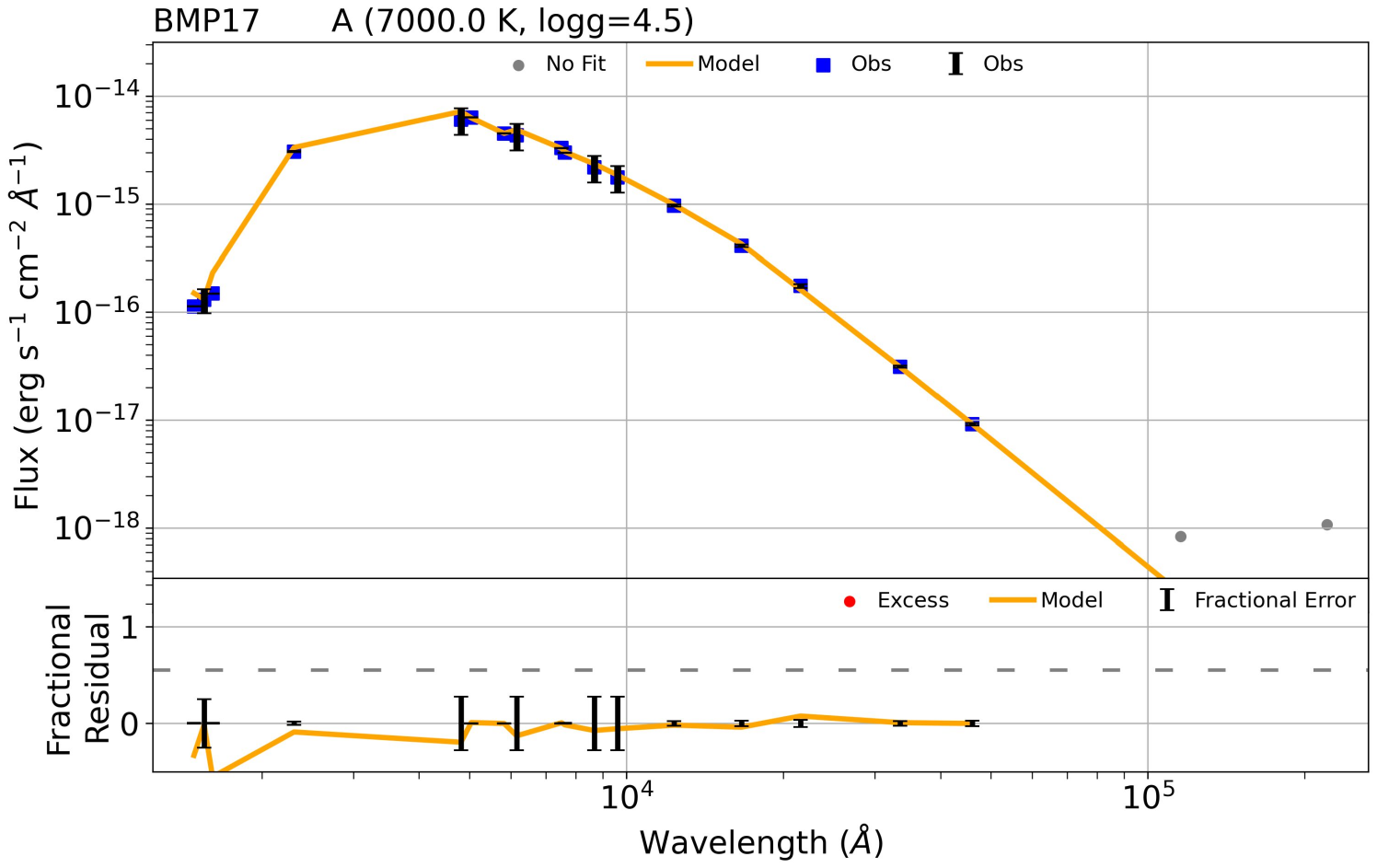} 
\includegraphics[scale=0.25]{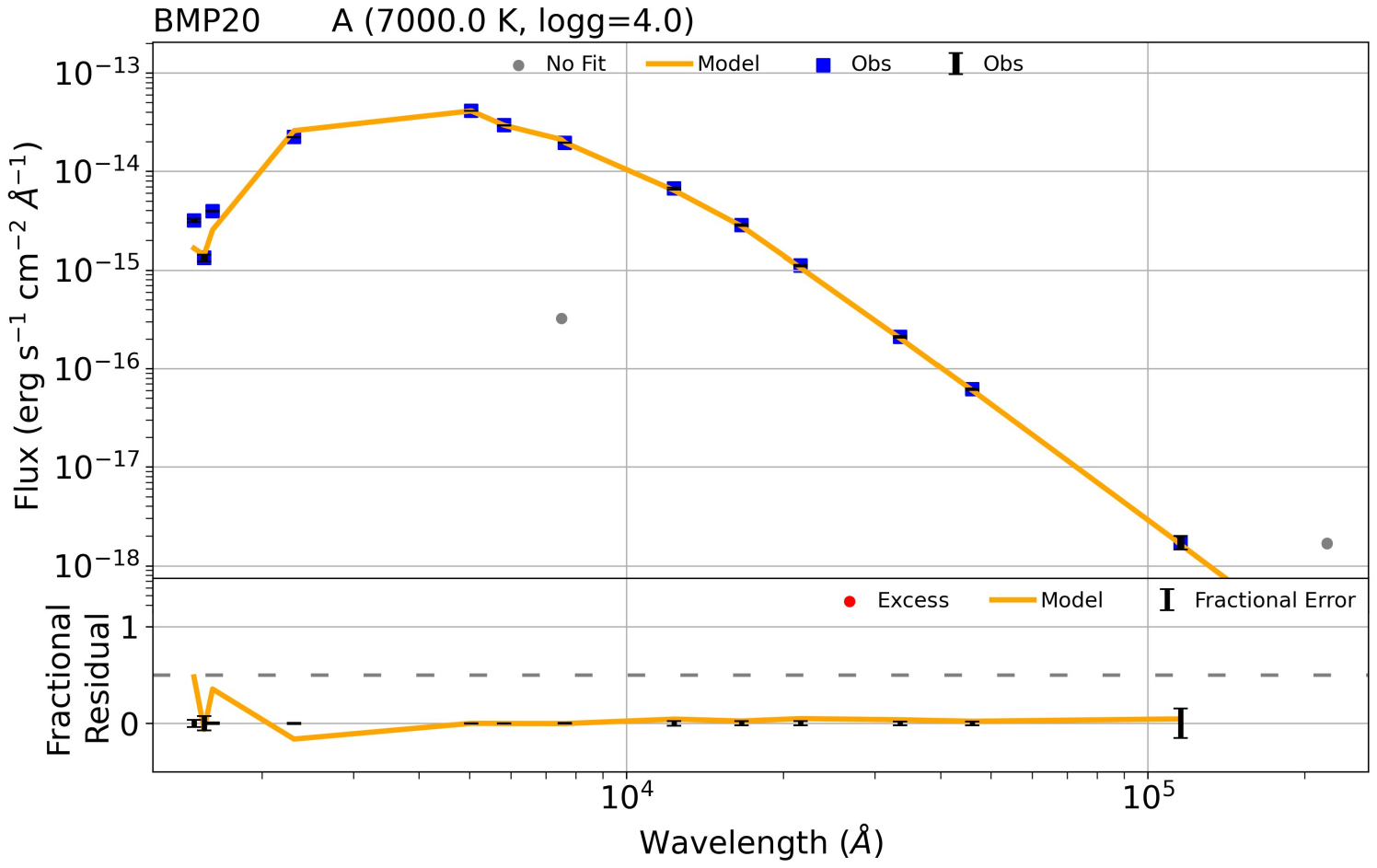}
\includegraphics[scale=0.25]{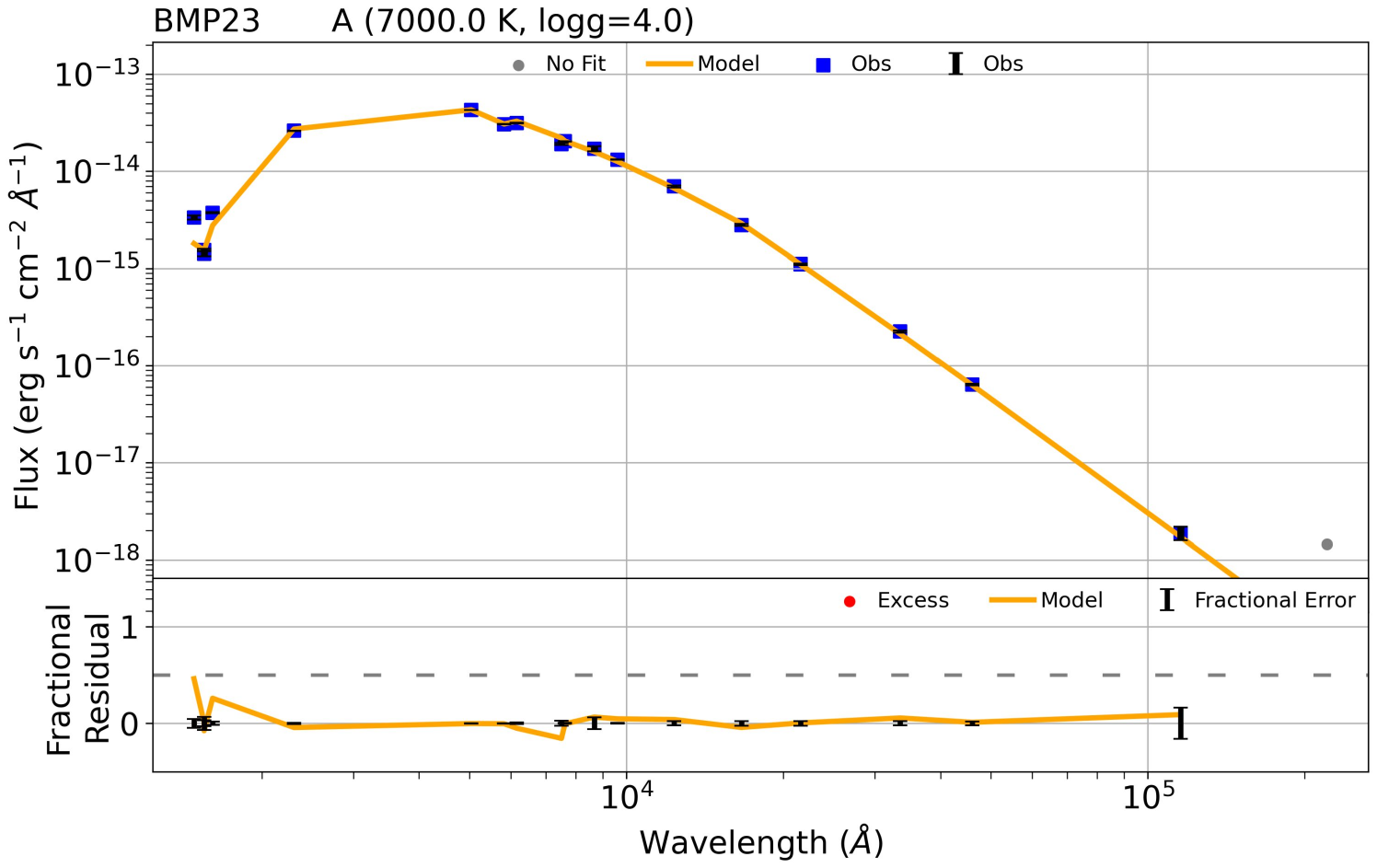}
\includegraphics[scale=0.25]{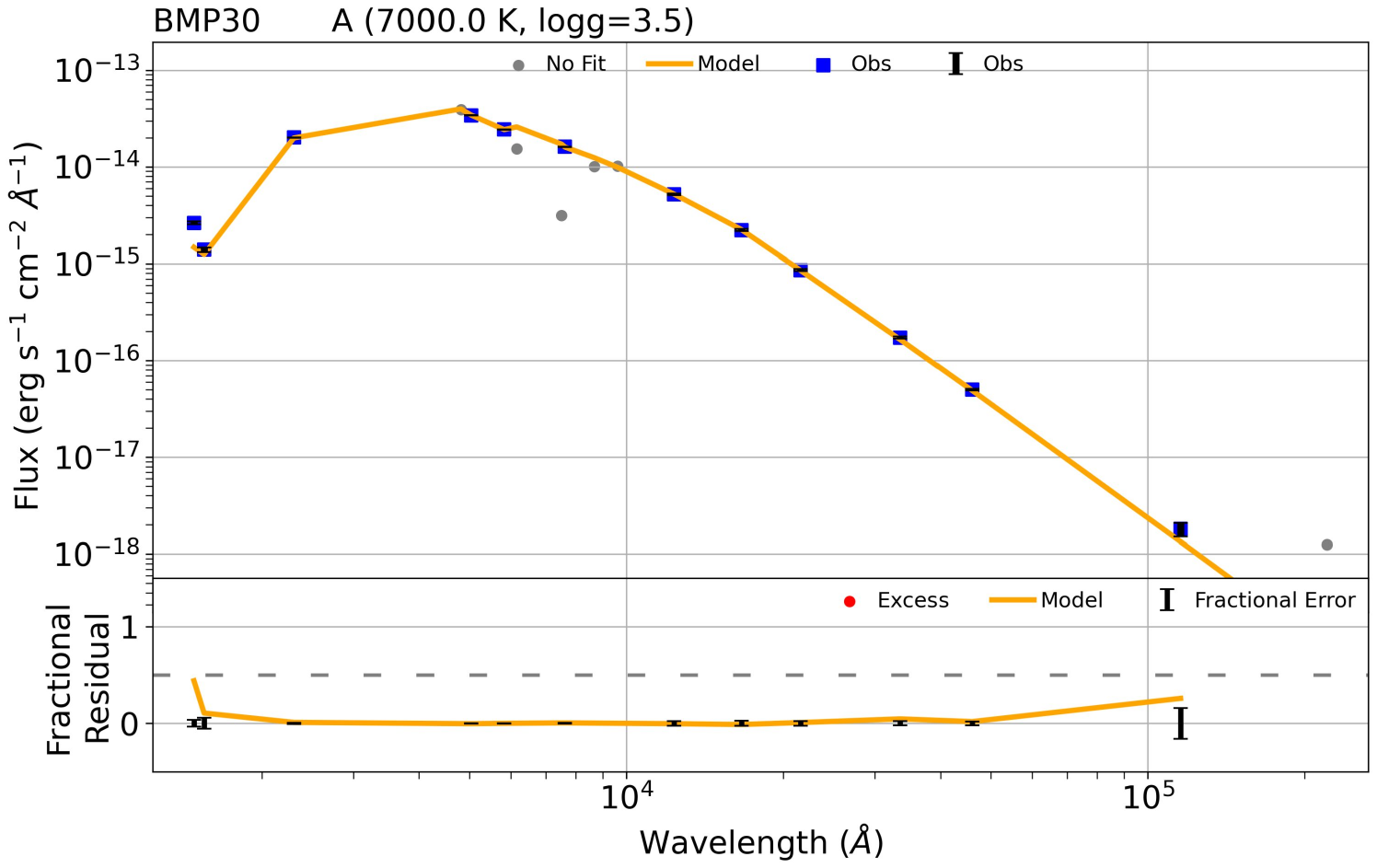} 
\includegraphics[scale=0.25]{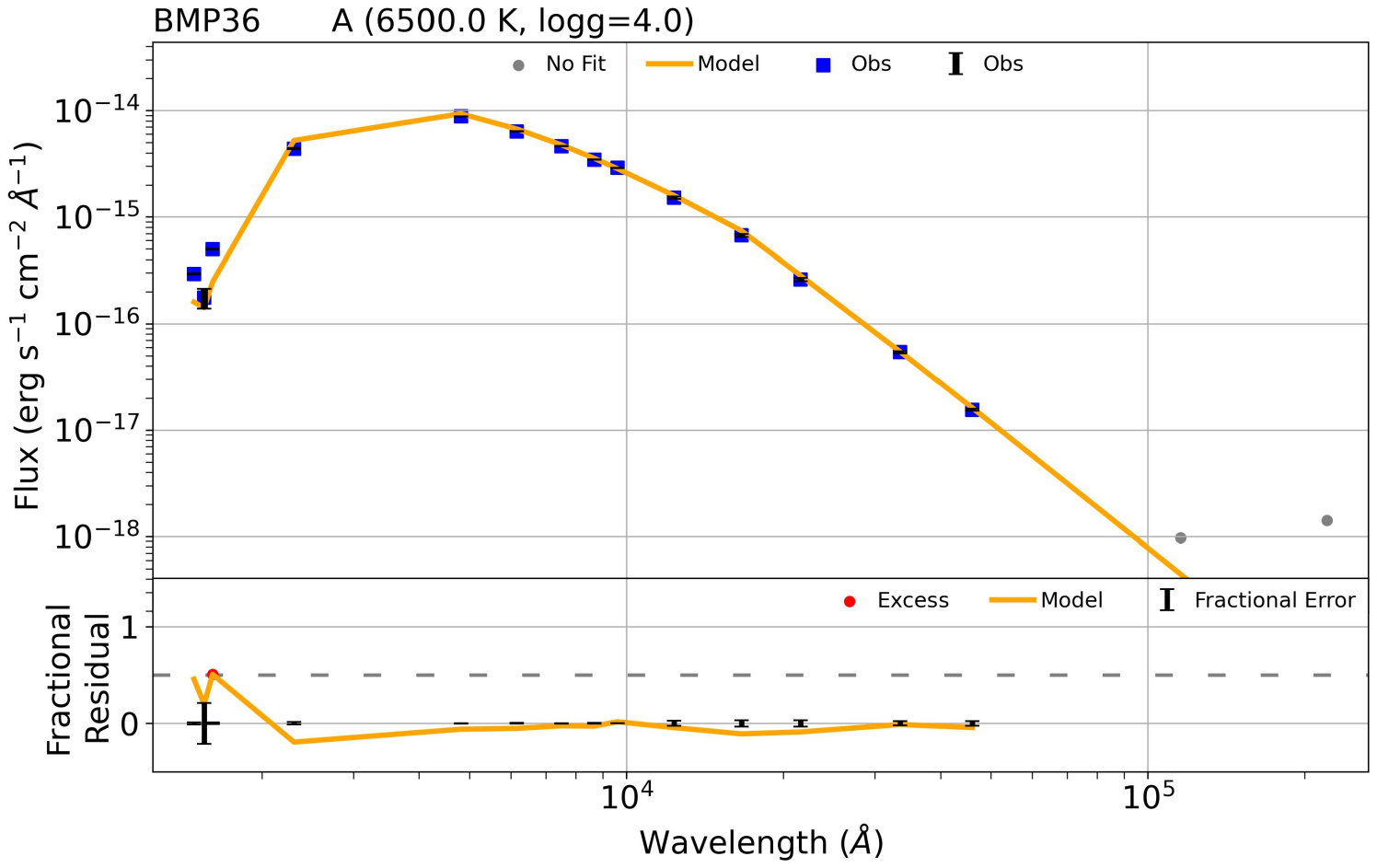} 
\includegraphics[scale=0.25]{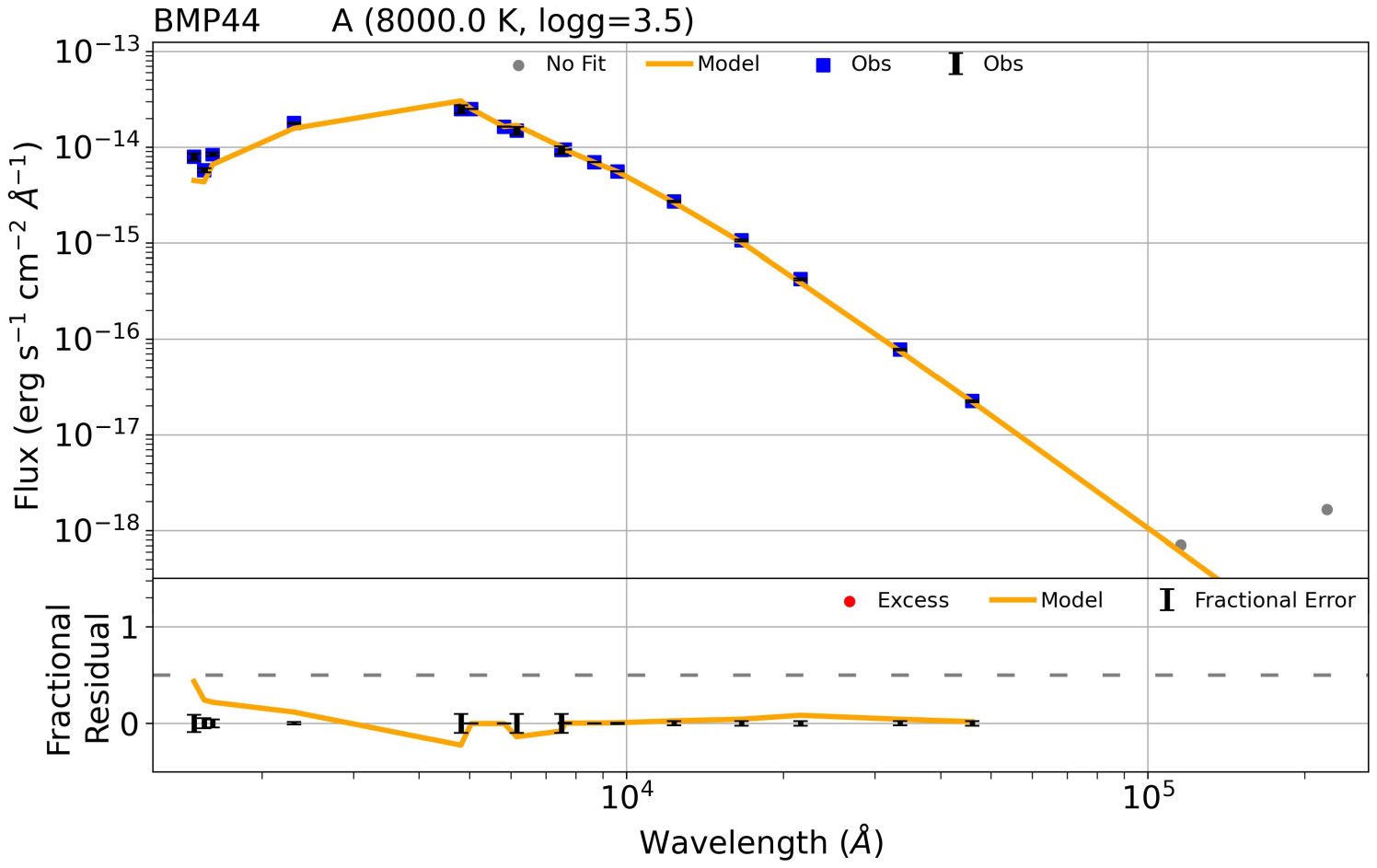} 
\includegraphics[scale=0.25]{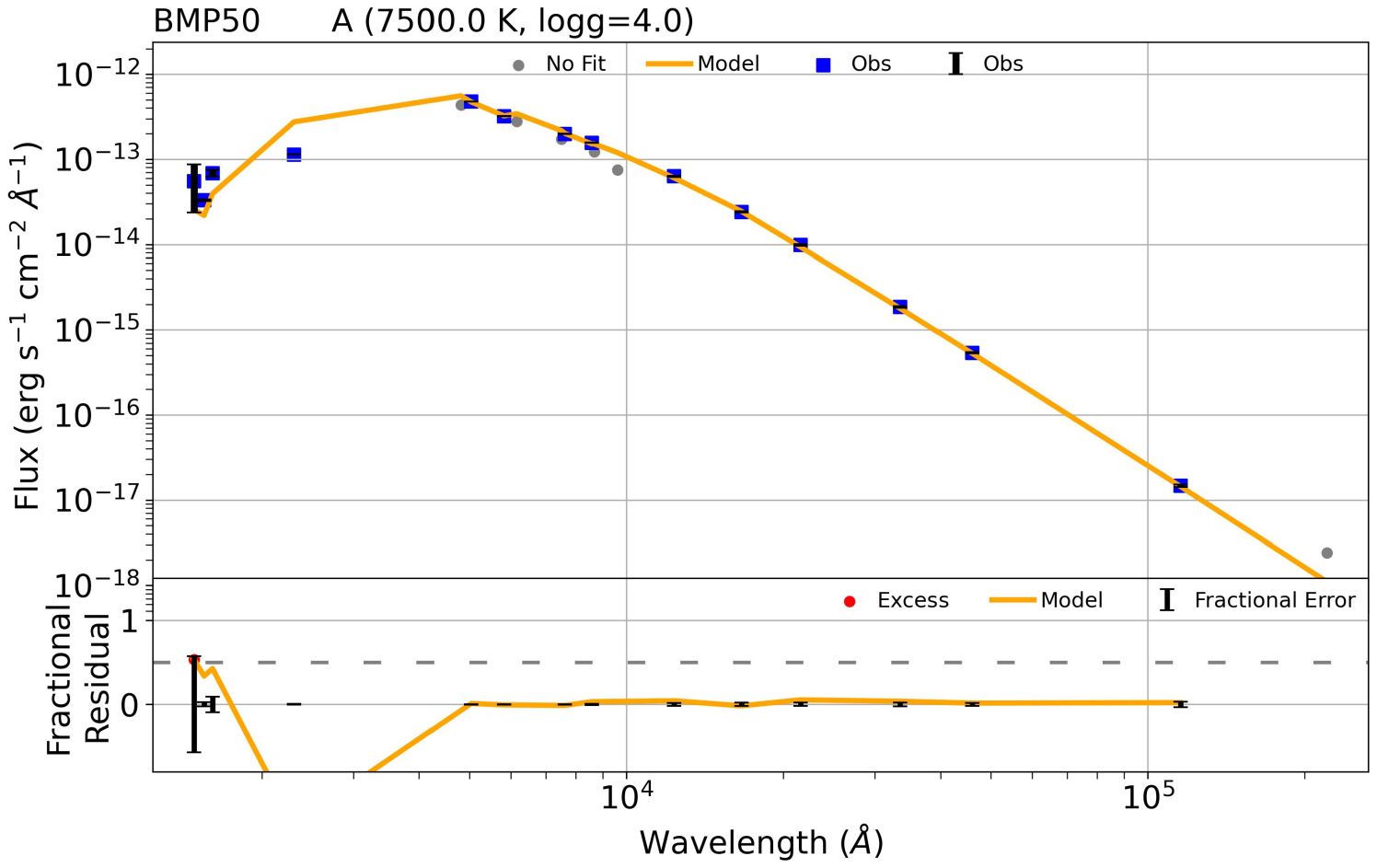}
 
\caption{The single component SED fits of BMP stars (cont.). }
\label{Fig. A1}
\end{figure*}

\begin{figure*}  

\includegraphics[scale=0.25]{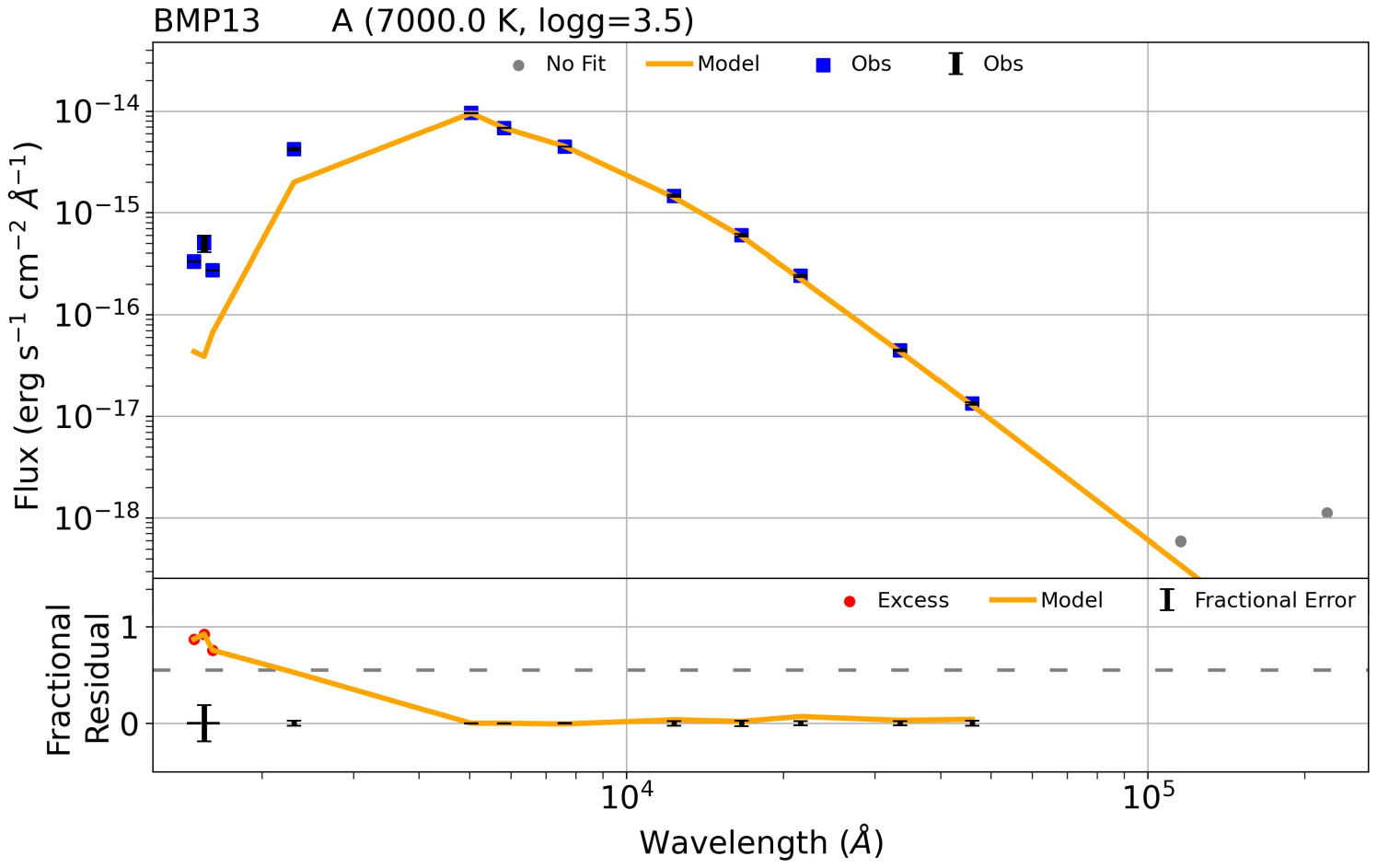}
\includegraphics[scale=0.25]{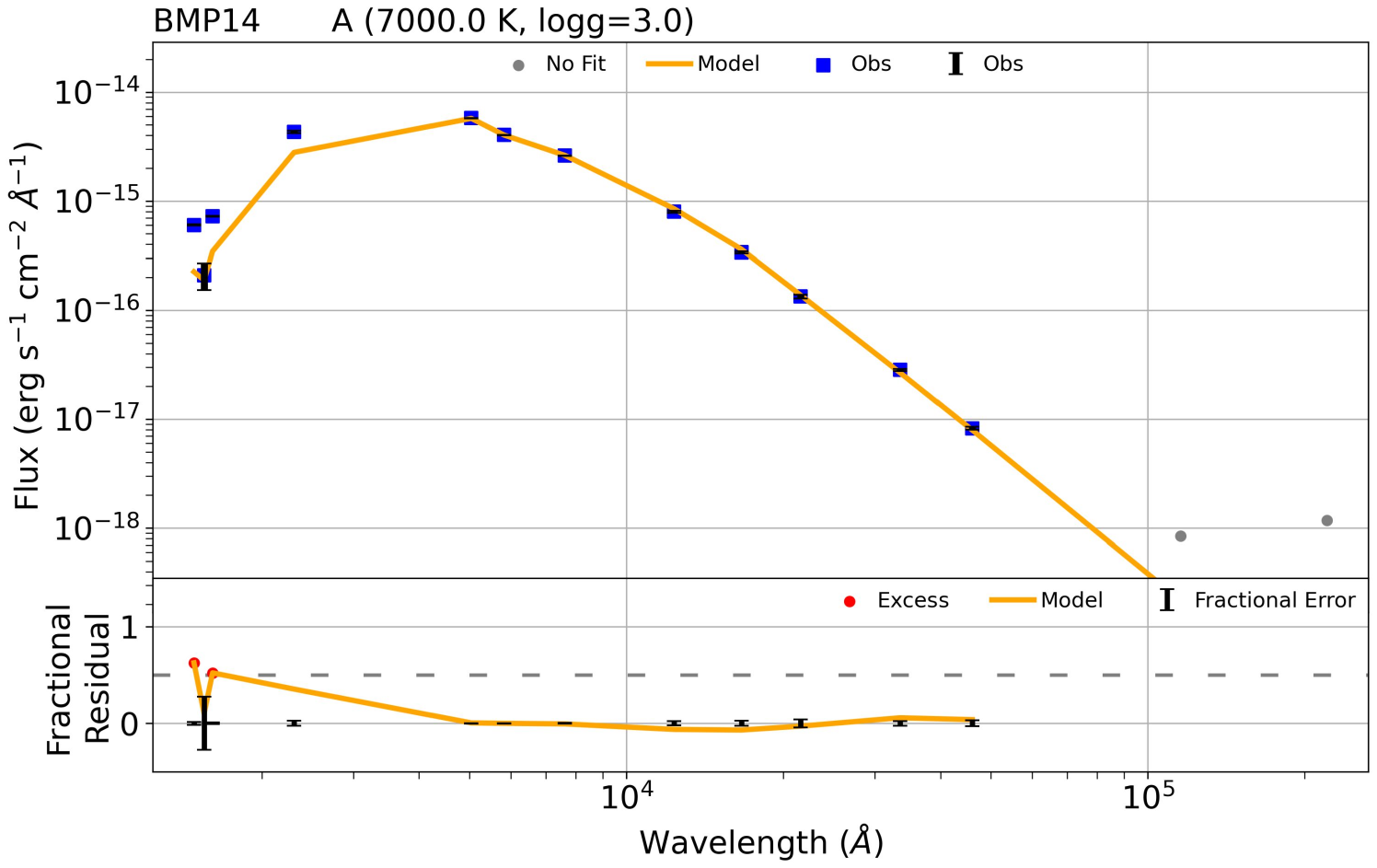}
\includegraphics[scale=0.25]{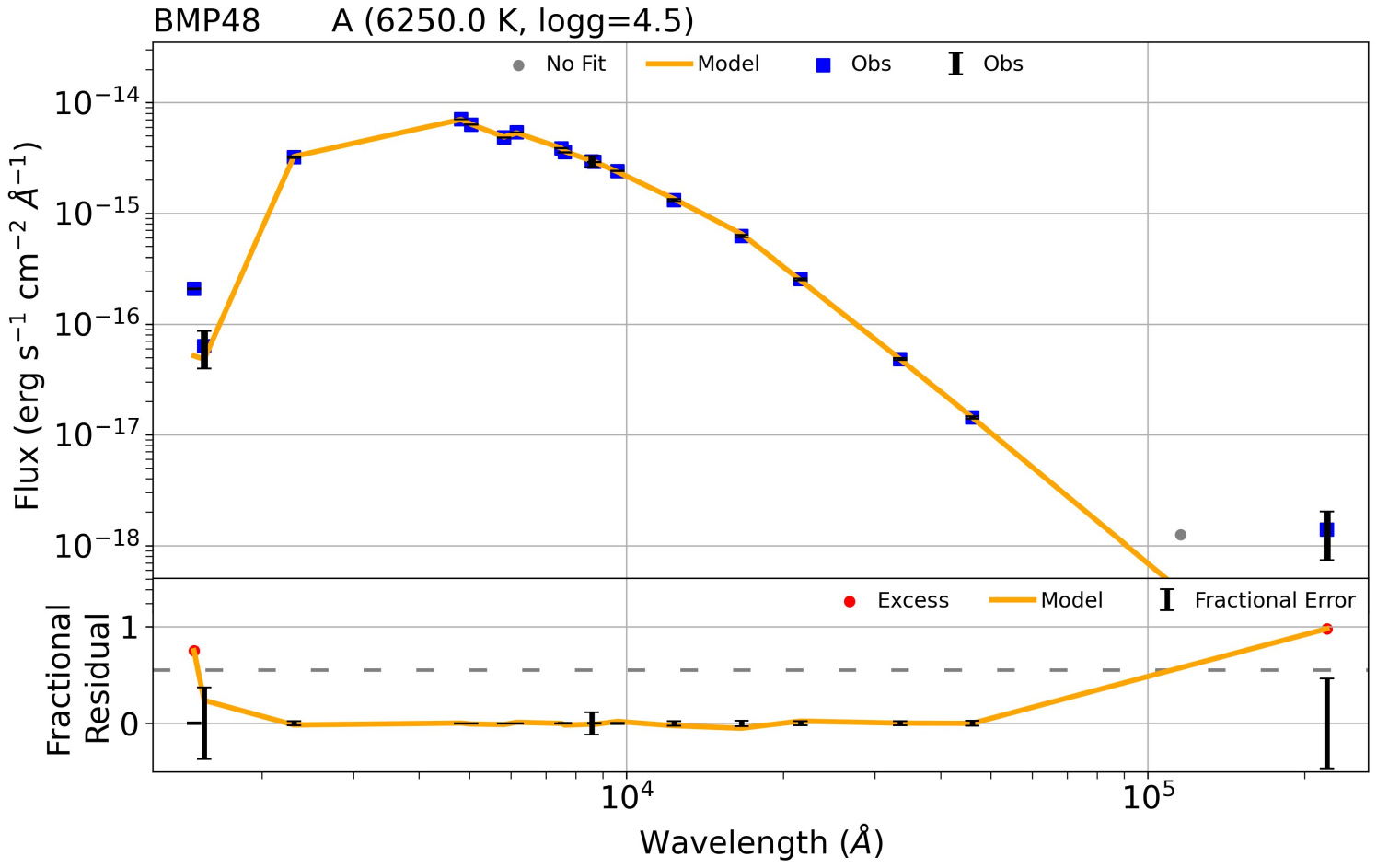}
 
\caption{The BMP stars showing UV excess but not fitted with the binary-component SEDs (cont.).} 
\label{Fig. A2}
\end{figure*}

\bsp
\label{lastpage}

\end{document}